\documentclass{aa}  

\usepackage{graphicx}
\usepackage{multirow}
\usepackage{txfonts}
\usepackage[colorlinks=true]{hyperref}
\begin{document}

   \title{A Magnetic Reconnection model for Hot Explosions in the Cool Atmosphere of the Sun}

   \author{Lei Ni\inst{1,4}
          \and
          Yajie Chen\inst{2,3}
          \and
          Hardi Peter\inst{3}
          \and
          Hui Tian\inst{2,5}
          \and
          Jun Lin\inst{1,4,6}
          }

   \institute{Yunnan Observatories, Chinese Academy of Sciences, Kunming, Yunnan 650216, P. R. China \\
              \email{leini@ynao.ac.cn}
         \and
             School of Earth and Space Sciences, Peking University, Beijing 100871, China \\
             \email{chenyajie@pku.edu.cn}
         \and
             Max Planck Institute for Solar System Research, Justus-von-Liebig-Weg 3, D-37077, G\"ottingen, Germany
         \and 
              Center for Astronomical Mega-Science, Chinese Academy of Sciences, 20A Datun Road, Chaoyang District, Beijing 100012, P. R. China
          \and
              Key Laboratory of Solar Activity, National Astronomical Observatories, Chinese Academy of Sciences, Beijing 100012, P. R. China.    
         \and
              University of Chinese Academy of Sciences, Beijing 100049, P. R. China.
         }

 %  \date{Received September 15, 1996; accepted March 16, 1997}

\abstract{
 \emph{Context.} UV bursts and Ellerman bombs are transient brightenings observed in the low solar atmospheres of emerging flux regions.  Magnetic reconnection is believed as the main mechanism to result in the formations of the two activities, which are usually formed far apart from each other. However, the observations also discovered the cospatial and cotemporal EBs and UV bursts, and their formation mechanisms are still not clear. The multi-thermal components with a large temperature span in these events challenge our understanding of magnetic reconnection and heating mechanisms in the partially ionized lower solar atmosphere.
  
 \emph{Aims.} We have studied magnetic reconnection between the emerging magnetic flux and back ground magnetic fields in the partially ionized and highly stratificated low solar atmosphere. We aim to explain the multi-thermal characteristics of UV bursts and find out whether EBs and UV burst can be generated in the same reconnection process, and how they are related with each other.  We also aim to unearth the important small-scale physics in these events.
  
 \emph{Methods.} The single-fluid MHD code NIRVANA was used to perform simulations. The background magnetic fields and emerging fields at the solar surface are reasonably strong. The initial plasma parameters are based on the C7 atmosphere model\citep{Avrett2008}. Cases with different resolutions have been simulated, and the effects of ambipolar diffusion, radiative cooling and heat conduction have been included. The current density, plasma density, temperature and velocity distributions in the main current sheet region have been analyzed. The Si~{\sc{iv}} emission spectrum has been synthesized.
  
  \emph{Results.} After the current sheet with dense photosphere plasma is emerged to $0.5$ \,Mm above the solar surface, plasmoid instability appears. The plasmoids collide and coalesce with each other, which makes the plasmas with different densities and temperatures mixed up in the turbulent reconnection region. Therefore, the hot plasmas corresponding to the UV emissions and colder plasmas corresponding to the emissions from other wavelenghts can move together and occur at about the same height. In the meantime, the hot turbulent structures basically concentrate above 0.4 Mm, whereas the cool plasmas extend to much lower heights to the bottom of the current sheet. These phenomena are consistent with the observations of \cite{Chen2019b}, in which UV bursts have a tendency to locate at the higher heights of corresponding EBs and all the EBs have partial overlap with corresponding UV burst in space. The synthesized Si~{\sc{iv}} line profiles are similar to the observed one in UV bursts, the enhanced wing of the line profiles can extend to about $100$\,km s$^{-1}$.  The differences are significant among the numerical results with different resolutions, which indicate that the realistic magnetic diffusivity is crucial to reveal the fine structures and realistic plasmas heating in these reconnection events. Our results also show that the reconnection heating contributed by ambipolar diffusion in the low chromosphere around the temperature minimum region is not efficient. }

 \keywords{  Magnetic reconnection; Magnetohydrodynamics (MHD); Magnetic fields ;Sun: chromosphere
               }
 \maketitle
%
%--------------------------------------------------------------------------------------------------------------------------------%

\section{Introduction}\label{sec:intro}
The low solar atmosphere is always in a dynamic state even in the so called quiet Sun region. Different scales of magnetic flux emergence happen all the time. The observational results indicate that the emerged magnetic fields can be as strong as several thousand Gauss in the photosphere\citep[e.g.,][]{Yan2017, Leenaarts2018, Getling2019, Liu2020, Yan2020}. Magnetic reconnection can happen between the emerging magnetic fields with opposite directions, or when the emerged ones interact with the back ground magnetic fields. The frequently observed compact transient brightenings, the bidirectional flows in these brights and the corresponding opposite magnetic fields approaching each other in the photosphere indicating these reconnection events. Numerous observations with high resolutions at different wavelengths have discovered different kinds of small scale reconnection events \citep[e.g.,][]{Xue2016, Zhao2017, Tian2018a, Huang2018, Huang2019, Yang2019} from the photosphere to the transition regions.

Ellerman bombs (EBs) \citep[e.g.,][]{Ellerman1917, Ding1998, Georgoulis2002, Watanabe2011, Nelson2015, Vissers2013, Yang2013, Vissers2015} are considered as one type of these low solar atmosphere reconnection events, which usually appear around the solar temperature minimum region (TMR). They show compact intense brightenings in images of the extended H${\alpha}$ wings, but no obvious signatures in H${\alpha}$ core images. The temperature increases in EBs are about $400$ to $3000$\,K \citep[e.g.,][]{Fang2006, Rutten2016, Hong2017}. The typical lifetime of a EB is about several minutes, and the typical size is about $1^{\prime \prime}$. 

In the past few years, the \emph{Interface Region Imaging Spectrograph} \citep[IRIS;][]{DePontieu2014} has identified a type of "hot bombs"\citep{Peter2014, Grubecka2016, Tian2016, Rouppe2017, Chitta2017, Tian2018a, Tian2018b, Gugliel2019, Chen2019a}, which are called UV bursts \citep[e.g.,][]{Young2018}. The observations indicate that UV bursts are also formed in magnetic reconnection process in the cool lower solar atmosphere. They share some similar characteristics with the traditional EBs (e.g., similar lifetimes and sizes). The strong Si~{\sc{iv}} emission, enhanced emission in the Mg~{\sc{ii}} wings but not core, Mn~{\sc{i}} absorption in the wings of Mg~{\sc{ii}}, deep absorption in Ni~{\sc{ii}} lines and compact brightenings in the AIA 1700 are usually observed in the UV bursts. Therefore, there are suggestions that some UV bursts might also be formed around the solar TMR region \citep[e.g.,][]{Peter2014, Tian2016} . However, the Si~{\sc{iv}} emissions in the UV bursts require that the temperature increase of about $2\times10^4$~K in the dense photosphere, or about $8\times10^4$~K in the upper chromosphere with much smaller density \citep{Rutten2016}. Though UV bursts have a signature in the UV continua at 1600 and 1700 observed by the \emph{Solar Dynamics Observatory} (\emph{SDO}) Atmospheric Imaging Assembly \citep[AIA;][]{Lemen2012}, but remain invisible in its He~{\sc{ii}} and higher-temperature coronal channels. The spectral profiles of Si~{\sc{iv}}, C~{\sc{ii}} and Mg~{\sc{ii}} h\&k emission lines in UV burst are significantly enhanced and broadened\citep[e.g.,][]{Peter2014, Tian2016}, the red and blue wings of the line profile of Si~{\sc{iv}} emission lines are usually about 100 km\,s$^{-1}$ away from line center. The commissioning observations with CHROMIS in Ca~{\sc{ii}} K uncover a whole new level of fine structures in UV bursts, which are highly dynamic blob-like substructure evolving on the timescale of seconds \cite{Rouppe2017}. The non-LTE inversions of the low-atmosphere reconnection based on the SST and IRIS observations also showed the blob-like substructures \cite{Vissers2019}. These blob-like fine structures are suggested as one of the mechanisms to cause the non-Gaussian broadened spectral profiles. However, we should also point out that the UV bursts with narrow line widths of $15-18$\,km\,s$^{-1}$ have also been observed \cite{Hou2016}. 

Recently, the relationship between EBs and UV bursts has been debated. The observations showed that the occurrence of cospatial and cotemporal EBs and UV bursts is between $10\%$ and $20\%$ \citep[e.g.,][]{Vissers2015, Tian2016, Chen2019b}, which suggests that the UV bursts  and their associated EBs are caused or modulated by a common physical process. However, \cite{Fang2017} pointed out that the observed H${\alpha}$ emission cannot be reproduced using non-local thermal equiplibrium (non-LTE) semi-empirical modeling if the temperature is above $10^4$~K. Therefore, the H${\alpha}$ emissions and Si~{\sc{iv}}emissions in these events should come from the plasmas with different temperatures. \cite{Chen2019b} have identified 161 EBs from the 1.6 m Goode Solar Telescope (GST), and ~20 of them reveal signatures of UV bursts in the IRIS images. They found that most of these UV bursts have a tendency to appear at the upper parts of their associated flame-like EBs. However, there are still about one third of the events, in which the formation heights of the UV bursts are not found to be higher than their corresponding EBs. 

The early numerical simulations about magnetic reconnection in the low solar atmosphere can qualitatively reproduce and explain the typical characteristics of the observed EBs \citep[e.g.,][]{Chen2001, Isobe2007, Archontis2009, Danilovic2017}. The large anomalous or numerical resistivities are usually applied to trigger the reconnection processes. The partial ionization effects are rarely considered in these simulations, and the reconnection magnetic fields are about dozens of Gauss, which makes the maximum temperature increase is only about several thousand Kelvin. \cite{Ni2015, Ni2016} have studied magnetic reconnection around the solar TMR with much stronger magnetic fields, the ambipolar diffusion effect resulted by the decoupling of ions and neutrals have been included and the extremely high resolutions make the magnetic diffusion in the reconnection process to be close to the realistic one. For the first time, the plasmoid cascading process has been shown and studied in the chromosphere magnetic reconnection by numerical simulations \citep{Ni2015}, the plasmas can be heated from several thousand Kelvin to above tens of thousands Kelvin by small scale shocks inside the plasmoids \citep{Ni2015, Ni2016} even around the solar temperature minimum region (TMR, $\sim 500$\,km above the solar surface). The multi-fluid simulations by including the iterations between ions and neutrals further prove that the plasmas indeed can be heated above tens of thousands Kelvin when the reconnection magnetic fields around the solar TMR  are strong \citep{Ni2018a, Ni2018b, Ni2018c}. Though the non-equilibrium ionization-recombination effect makes the temperature increase in the reconnection process more difficult \citep{Ni2018a}. 

The plasmoids in the reconnection current sheets are suggested to correspond to the blob like structures in the observations. The recent simulation results show that the plasmoid instability in the reconnection region can significantly broaden the Si~{\sc{iv}} spectral line profiles \citep{Innes2015, Rouppe2017}, the strong enhancement of line cores along with the increased emission in the line wings can both be produced when plasmoid instability occurs.

 The flux emergence process has been frequently modeled and applied to simulate magnetic reconnection events (e.g., flares, jets and surges) with different length scales from solar chromosphere to corona \citep[e.g.,][]{Heyvaerts1977, Shibata1992, Yokoyama1995, Galsgaard2005, Archontis2005, Takasao2013, Nobrega2016, Nobrega2018}. When the resolutions are high enough and the Lundquist number exceeds the critical value\citep[e.g.,][]{Leake2013, Ni2015, Murphy2015}, the plasmoids can always be identified  in the interaction regions where the magnetic fields with opposite directions meet with each other even in the chromosphere. The recent flux emergence simulations \citep{Nobrega2016, Rouppe2017, Nobrega2018} with a length scale of 10 Mm have shown the formations of plasmoids above the up chromosphere (1.5 Mm above the solar surface). 

\cite{Hansteen2017} has shown that magnetic reconnection between the emerged magnetic fields can cause  the formations of EBs in the photopshere, UV bursts in the middle chromosphere, respectively. However, the EBs and UV bursts appear at different reconnection process, no cospatial or cotemporal  EBs and UV burtsts were produced. The same three dimensional (3D) radiative MHD code was then used to study the UV bursts which connect with EBs in \cite{Hansteen2019}. Their numerical results show that a long-lasting current sheet that extends over different scale heights through the low solar atmosphere is formed. The part above 1 Mm of such a long current sheet is very hot and the temperature reaches about $10^5$\,K, the temperature of the lower part is below $10^4$\,K. The plasmoid like fine structures are not mentioned and shown in their work.Their simulation results indicate that EBs and UV bursts are occasionally formed at opposite ends of a long current sheet \citep{Hansteen2019}, and this can be one of the possible models to explain the UV bursts which connect with the EBs.   

In this work, magnetic reconnection resulted by flux emergence in the stratificated low solar atmosphere has been numerically studied. We have proposed a model to explain the UV bursts with multi-thermal characteristics and the UV bursts which are relating with EBs. The simulations have been performed by using different resolutions, the effects of radiative cooling, ambipolar diffusion and heat conduction have also been discussed. The remaining part of this paper is structured as follows. Section 2 describes the numerical models, initial and boundary conditions. Section 3 presents our numerical results. A summary and discussions are presented in section 4.    
      
%--------------------------------------------------------------------------------------------------------------------------------%
\section{Numerical setup} \label{sec:model}

\subsection{Model equations} \label{sec:equations}

We performed 2.5D MHD simulations in Cartesian geometry using the single-fluid MHD code NIRVANA \citep[version 3.6;][]{Ziegler2011}. The solved MHD equations are as follows:

\begin{eqnarray}
\frac{\partial \rho }{\partial t}&=&-\nabla \cdot \left (\rho \mathbf{v}\right ) \\
\frac{\partial \left ( \rho \mathbf{v} \right )}{\partial t} &=&- \nabla \cdot \left [ \rho \mathbf{vv} +\left ( p+\frac{1}{2\mu _{0}}\left | \mathbf{B} \right |^{2} \right )I-\frac{1}{\mu _{0}}\mathbf{BB} \right ] \nonumber \\
&&+\nabla \cdot \tau +\rho \mathbf{g} \\
\frac{\partial e}{\partial t}&=&-\nabla \cdot\left [ \left ( e+p+\frac{1}{2\mu _{0}}\left | \mathbf{B} \right |^{2} \right )\mathbf{v} \right ]  \nonumber \\
&&+\nabla \cdot\left [ \frac{1}{\mu _{0}}\left ( \mathbf{v}\cdot \mathbf{B} \right )\mathbf{B}   \right ] \nonumber \\
&&+\nabla \cdot \left [ \mathbf{v}\tau +\frac{\eta }{\mu _{0}}\mathbf{B}\times \left ( \nabla \times \mathbf{B} \right )\right ]       \nonumber \\
&&-\nabla \cdot \left [ \frac{1 }{\mu _{0}}\mathbf{B}\times\mathbf{E}_{AD}+\mathbf{F}_{c}\right ] \nonumber \\
&&+\rho \mathbf{g}\cdot \mathbf{v}+L_{rad}+H  \label{enereq} \\
\frac{\partial \mathbf{B}}{\partial t} &=& \nabla \times \left ( \mathbf{v}\times \mathbf{B}-\eta \nabla \times \mathbf{B}+\mathbf{E}_{AD} \right )  \label{indeq} \\
e&=&\frac{p}{\gamma -1}+\frac{1}{2}\rho \left | \mathbf{v} \right |^{2}+\frac{1}{2\mu _{0}}\left | \mathbf{B} \right |^{2} \\
p&=&\frac{\left ( 1+Y_{i} \right )\rho}{m_{i}}k_{B}T
\end{eqnarray}
where $\rho$, $\mathbf{v}, \mathbf{B}, p, T, e, Y_{i}$ are mass density, fluid velocity, magnetic field, thermal pressure, temperature, total energy density, and ionization fraction of the plasma, respectively. The gravitational acceleration of the sun is $\mathbf{g}$=$-$273.9 m s$^{-2}$ $\mathbf{e}_{y}$, and $m_{i}$ is the mass of proton. We set the ratio of specific heats as $\gamma = 5/3$. Dynamic viscosity was included, and $\tau=\nu \left [ \nabla \mathbf{v}+\left ( \nabla \mathbf{v} \right )^{\mathrm{T}}-\frac{2}{3}\left ( \nabla\cdot  \mathbf{v}  \right )I \right ]$ is the stress tensor, where $\nu$ is the dynamic viscosity coefficient and its unit is kg m$^{-1}$ s$^{-1}$. 

We applied the physical magnetic diffusion in our simulations. Similar to \citet{Khomenko2012}, collisions between electrons and ions and those between neutrals and electrons both contribute to the magnetic diffusivity $\eta$, which was given by:
\begin{equation}
\eta=\frac{m_{e} \left( \nu_{ei}+ \nu_{en} \right) }{e^{2}n_{e}\mu_{0}}
\label{eta_org}
\end{equation}
where $\nu_{ei}$ and $\nu_{en}$ are collisional frequencies between electrons and ions and those between electrons and neutrals. According to \citet{Spitzer1962} and \citet{Braginskii1965}, Equation~\ref{eta_org} can be simplified as:

\begin{equation}
\eta=10^{9} T^{-\frac{3}{2}}+1.762\times 10^{-3} \left( \frac{1}{Y_{i}}-1 \right) \sqrt{T} 
\label{etaeq}
\end{equation}

We also used the classic form of anisotropic heat conduction flux:
\begin{equation}
\mathbf{F}_{c}=-\kappa _{\parallel }\left ( \nabla T\cdot \hat{\mathbf{B}} \right )\hat{\mathbf{B}}-\kappa _{\perp}\left [ \nabla T-\left ( \nabla T\cdot \hat{\mathbf{B}} \right )\hat{\mathbf{B}} \right ]
\end{equation}
where $\hat{\mathbf{B}}=\mathbf{B}/\left | \mathbf{B} \right |$ is the unit vector in the direction of magnetic field. According to \citet{Orrall1961} and \citet{Ni2016}, the parallel and perpendicular conductivity coefficients, $\kappa _{\parallel }$ and $\kappa _{\perp}$, was give by:
 \begin{eqnarray}
 \kappa_{\parallel}&=& 1.96 \times 10^{-11} T^{\frac{32}{13}} + 10^{8}T^{-2.5} \\
 \kappa_{\perp}&=& 10^{8}T^{-2.5} 
 \end{eqnarray}
in which $10^{8}T^{-2.5}$ and $1.96 \times 10^{-11} T^{\frac{32}{13}}$ are the terms associated with neutral and ionized particles, respectively.

Radiative cooling function $L_{rad}$ was taken from \citet{Gan1990}:
\begin{equation}
L_{rad}=-1.547\times 10^{-42} Y_{i} \left( \frac{\rho}{m_{i}}\right)^{2} \alpha T^{1.5}  
\end{equation}
and the heating function $H$ was the same as that in \citet{Ni2016}:
\begin{equation}
H=1.547\times 10^{-42} Y_{i}  \frac{\rho \rho_{0}}{m_{i}^{2}} \alpha T^{1.5}_{0}  
\end{equation}
where $\rho_{0}$ and $T_{0}$ are the initial mass density and temperature.

The ambipolar diffusion field $\mathbf{E}_{AD}$ in the energy equation (\ref{enereq}) and induction equation (\ref{indeq}) was given by:
\begin{equation}
\mathbf{E}_{AD}=\frac{1}{\mu_{0}}\eta _{AD}\left [ \left ( \nabla \times \mathbf{B} \right ) \times \mathbf{B}  \right ] \times \mathbf{B} 
\end{equation}
where $\eta _{AD}$ is the ambipolar diffusion coefficient. We took the same $\eta _{AD}$ given in \citet{Ni2015}:
\begin{equation}
\eta _{AD}=1.65\times 10^{-11}\left ( \frac{1}{Y_{i}}-1 \right )\frac{1}{\rho ^{2}\sqrt{T}} \label{eqetaad}
\end{equation}

\subsection{Initial conditions} \label{sec:incon}

\begin{figure}
\centering
\includegraphics[width=\hsize]{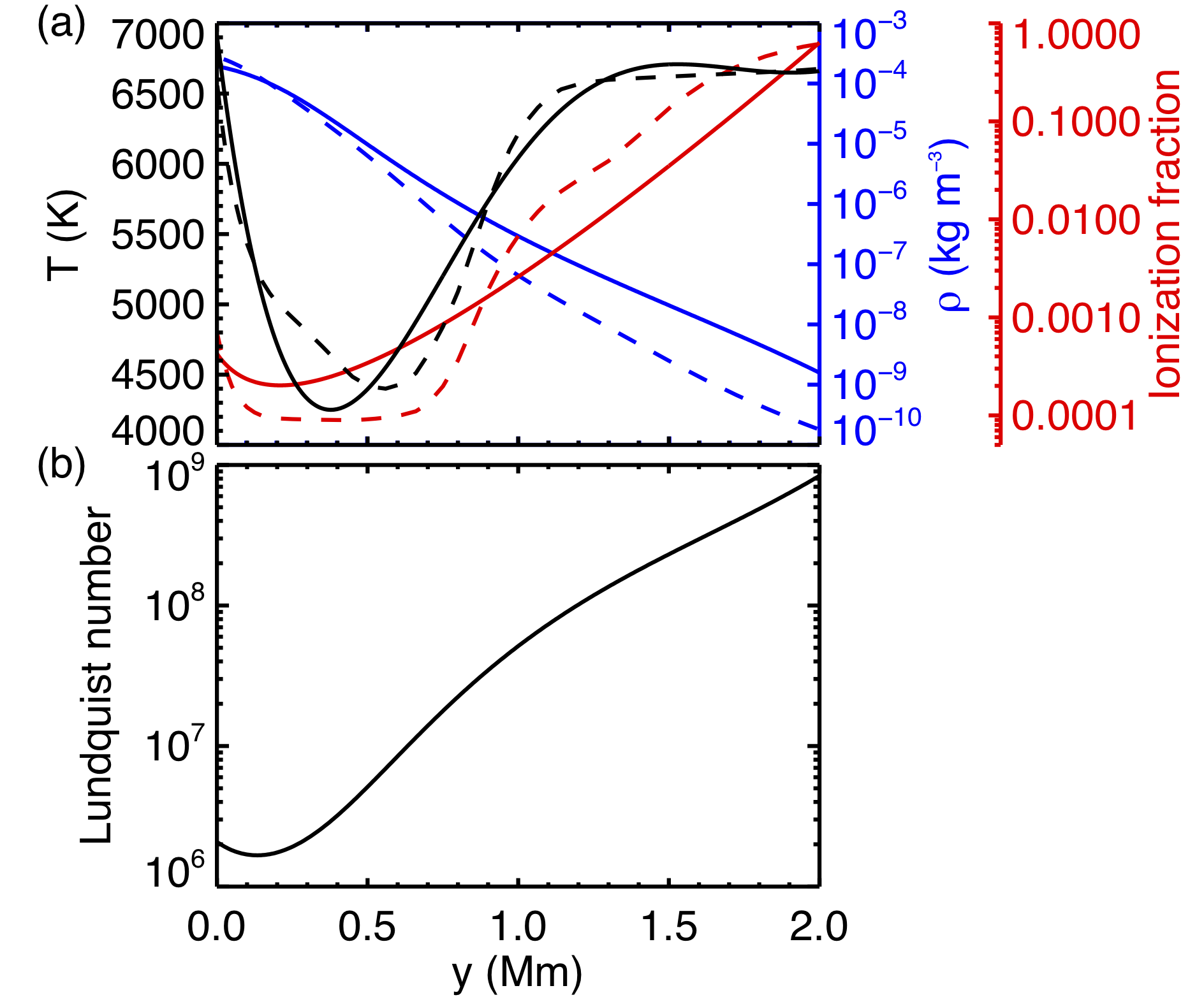}
\caption{The initial state of the simulations. (a) Temperature (black), density (blue), and ionization fraction (red) of the atmosphere. (b) The initial Lundquist number varies with the height. The corresponding  dashed lines represent the exact values in C7 model.  
}\label{f1}
\end{figure}

The simulation domain extends from $-$8 to 8 Mm in the horizontal (x) direction and from 0 to 2 Mm in the vertical (y) direction, and the photosphere is at $y=0$. We applied adaptive mesh refinement (AMR) in our simulations, and the simulations started from a base-level grid of $768 \times 384$. We used different highest refinement levels of 1, 3, 5, and 7 in cases 1, 2, 3, and 4 to investigate the effects of increasing spatial resolution. The minimum grid size is about 45 m when the highest refinement level reaches 7. The output raw data with non-uniform grids  based on the AMR skill are converted to uniform IDL data, which are then used to plot the figures in this work. The IDL data with different grid sizes can be chosen according to our choices.

Based on the C7 atmosphere model \citep{Avrett2008}, the function of the initial temperature profile is as below:
\begin{eqnarray}
T&=&-1228 \left ( \frac{y}{L_{0}} \right) ^{5}+9749 \left ( \frac{y}{L_{0}} \right) ^{4}-2.843\times10^{4}\left ( \frac{y}{L_{0}} \right) ^{3}  \nonumber  \\
&&+3.621\times10^{4}\left ( \frac{y}{L_{0}} \right) ^{2}-1.717\times10^{4}\left ( \frac{y}{L_{0}} \right) +6912 
\end{eqnarray}
where $L_{0}=10^{6}$ m, y is in the unit m, and T is in the unit K. We set the ionization fraction $(Y_{i})$ as:
\begin{equation}
Y_{i}=3 \times 10^{-5}  \frac {\exp \left( 3.486\times y_{1}^{1.2}  + 2.4\times y_{1}^{0.9} + y_{1}^{0.3} \right) }{17\times y_{1}^{2.2}+2.639\times y_{1}^{1.8}+0.006}
\end{equation}
where $y_{1}=y/L_{0} +0.12$. It is worth mentioning that the $Y_{i}$ does not change with time in our simulations. We chose the mass density at the photosphere in the C7 model as the mass density at our bottom boundary, and then we numerically solved the hydrostatic equilibrium equation $\nabla p=-\rho g$ to obtain the initial mass density in our simulations. The initial temperature, ionization fraction, and mass density of the atmosphere are presented in Figure~\ref{f1}. We mimicked a temperature minimum region at the heights around 400 km, i.e. the temperature drops from $\sim 6900$ K to $\sim 4200$ K and then rise to $\sim6600$ K. The ionization fraction varies from $\sim10^{-4}$ to $\sim 0.6$, and the resulting mass density drops from  $\sim 10^{-4}$~kg~m$^{-3}$ at the bottom boundary to $\sim10^{-9}$~kg~m$^{-3}$ at the top boundary.

We also assumed a uniform initial magnetic field:
\begin{eqnarray}
B_{x0}&=&-0.85 b_{0}  \\
B_{y0}&=&-0.35 b_{0}  \\
B_{z0}&=&c_{0}b_{0}
\end{eqnarray}
where $c_{0}$ and $b_{0}$ are constants. We chose $c_{0}=0.5$ and $b_{0}=600$~G for cases 5, 5a, 5b, and $c_{0}=1$ and $b_{0}=500$~G for other cases.  

Based on the initial temperature and ionization fraction of our simulations, we found that the initial $\eta$ is in the order of $10^{4}$ $m^{2}$ $s^{-1}$ in the whole simulation domains. Then we estimated the corresponding Lundquist number $Lv_{A}/\eta$, where $L=1~Mm$ is the typical length scale of the system and $v_{A}=B/\sqrt{\mu_{0}\rho}$ is the Alfv\'en speed. The initial Lundquist number is also shown in Figure~\ref{f1}, and it has values from $\sim10^{6}$ to $\sim10^{9}$. These values significantly exceed the critical Lundquist number ($10^{3} \sim 10^{4}$) for the onset of plasmoid instability \citep[e.g.,][]{Bhattacharjee2009, Huang2009, Ni2013, Leake2012, Leake2013}.

\subsection{Boundary conditions} \label{sec:bdcon}

We used outflow boundaries at the top boundary. The fluid is only allowed to flow out of the computation domain by assuming:
\begin{displaymath}
  v_{yug} = \left\{ \begin{array}{ll}
     v_{yul} & \textrm{ if $v_{yul}>0$ } \\
     -v_{yul} & \textrm{if $v_{yul}<0$},
     \end{array}  \right.
  \end{displaymath}   
where $v_{yug}$ and $v_{yul}$ separately represent the velocities at the two top ghost layers and at the last two layers inside the simulation box in the $y$-direction. The gradients of velocities in the $x$-direction and $z$-direction vanish at the top boundary by assuming $\frac{\partial v_x}{\partial y}=0$, $\frac{\partial v_z}{\partial y}=0$. Similarly, the gradients of the thermal energy, mass density and parallel components of magnetic field are set to zero. The perpendicular component of the magnetic field is obtained by divergence--free extrapolation of the magnetic field. We set the dynamic viscosity coefficient $\nu = 10^4 \rho [1+10(1+tanh\frac{y-1.95L_0}{0.1L_0})]$ to make an enhancement of viscous diffusion around the top boundary.

To drive the reconnection, we mimicked a flux emergence process similar to \citet{Ni2017} by setting the magnetic fields at the bottom boundary as:
\begin{eqnarray}
B_{xb}&=&\left\{\begin{matrix}
-0.85b_{0}+b_{1}\frac{\left ( y-y_{0} \right )L_{0}^{1.6}t}{\left [ x^{2} +( \left ( y-y_{0} \right )^{2} \right ]^{1.3}t_{0}} &,&~t\leqslant t_{0}\\ 
-0.85b_{0}+b_{1}\frac{\left ( y-y_{0} \right )L_{0}^{1.6}}{\left [ x^{2} +( \left ( y-y_{0} \right )^{2} \right ]^{1.3}}  &,&~t> t_{0}
\end{matrix}\right.
\\
B_{yb}&=&\left\{\begin{matrix}
-0.35b_{0}-b_{1}\frac{xL_{0}^{1.6}t}{\left [ x^{2} +( \left ( y-y_{0} \right )^{2} \right ]^{1.3}t_{0}} &,&~t\leqslant t_{0}\\ 
-0.35b_{0}-b_{1}\frac{xL_{0}^{1.6}}{\left [ x^{2} +( \left ( y-y_{0} \right )^{2} \right ]^{1.3}}  &,&~t> t_{0}
\end{matrix}\right.
\\
B_{zb}&=&c_{0}b_{0}
\end{eqnarray}
where $t_{0}=$ 60 s. The emerging flux and the highest heights it can reach increase monotonically with $y_{0}$ and $b_{1}$. The divergence-free condition of the magnetic field in the simulation domain is satisfied. We chose $y_{0}=-3\times10^{5}$ m and $b_{1}=$ 500 G for all the cases except Cases 5, 5a and 5b. We set $y_{0}=-2\times10^{5}$ m and $b_{1}=$ 600 G in Cases 5, 5a and 5b so that the emerging flux reaches higher heights. The emerging magnetic field with such a form makes the reconnection magnetic fields in $(x,y)$ plane around the temperature minimum region to be about several hundred Gauss. The small scale flux emergences with a maximum strength of several hundred Gauss are frequently observed in reconnection events of UV bursts and EBs \citep[e.g.,][]{Wang2020, Chen2019a}, the much stronger magnetic fields of over several thousand Gauss have also been observed in active regions\citep[e.g.,][]{Getling2019, Liu2020, Leenaarts2018, Yan2017, Yan2020}. Therefore, the strength of reconnection magnetic fields in our simulations are reasonable.

The fluid velocity was set to zero, the density and thermal energy was fixed with the initial values at the bottom boundary. In addition, periodic boundary conditions were used in the horizontal direction.

%--------------------------------------------------------------------------------------------------------------------------------%
\section{Results and discussions} \label{sec:results}

\begin{table*}[]
\centering
\caption{Differences among simulation cases.}
\label{tab1}
\begin{tabular}{l|c|ccc|cccc}
\hline
\multirow{2}{*}{Models} & \multirow{2}{*}{\begin{tabular}[c]{@{}c@{}}Highest\\ AMR Levels\end{tabular}} & \multirow{2}{*}{\begin{tabular}[c]{@{}c@{}}Radiative\\ Cooling\end{tabular}} & \multirow{2}{*}{\begin{tabular}[c]{@{}c@{}}Heat \\ Conduction\end{tabular}} & \multirow{2}{*}{\begin{tabular}[c]{@{}c@{}}Ambipolar \\ Diffusion\end{tabular}} & \multicolumn{4}{c}{Parameters} \\ \cline{6-9} 
 &  &  &  &  & $c_{0}$ & $b_{0}$ (G) & $b_{1}$ (G) & $y_{0}$ ($10^{5}$ m) \\ \hline
1 & 1 & Yes & Yes & Yes & 1 & 500 & 500 & -3 \\ \hline
2 & 3 & Yes & Yes & Yes & 1 & 500 & 500 & -3 \\ \hline
3 & 5 & Yes & Yes & Yes & 1 & 500 & 500 & -3 \\ \hline
3a & 5 & No & Yes & Yes & 1 & 500 & 500 & -3 \\ \hline
3b & 5 & Yes & No & Yes & 1 & 500 & 500 & -3 \\ \hline
3c & 5 & Yes & Yes & No & 1 & 500 & 500 & -3 \\ \hline
4 & 7 & Yes & Yes & Yes & 1 & 500 & 500 & -3 \\ \hline
5 & 5 & Yes & No & No & 0.5 & 600 & 600 & -2 \\ \hline
5a & 3 & Yes & No & No & 0.5 & 600 & 600 & -2 \\ \hline
5b & 3 & Yes & No & Yes & 0.5 & 600 & 600 & -2 \\ \hline
\end{tabular}%
\end{table*}

We have tested 10 different cases. As we mentioned before, we used the highest AMR level of 1, 3, 5, 7 in Cases 1, 2, 3, 4, respectively. Radiative cooling, heat conduction, and ambipolar diffusion are all included in these cases. Effects of increasing spatial resolutions are studied based on these cases. 

Then we turned off the radiative cooling, heat conduction, and ambipolar diffusion in Cases 3a, 3b, and 3c, respectively, to investigate the roles of different terms in our simulations, the highest refinement level in these cases are 5. We also ran cases (Case 5, 5a, 5b) with larger emerging flux. In Case 5 and 5a, the highest AMR refinement level is 5 and 3 respectively, but excluding ambipolar diffusion and heat conduction. The highest AMR refinement level in Case 5b is also 3, but the ambipolar diffusion is included.  A summary of our simulation models is shown in Table~\ref{tab1}.

%-----------------------------------------------------------------
\subsection{Simulations with different spatial resolution} \label{sec:resolution}

\begin{figure*} 
\centering {\includegraphics[width=\textwidth]{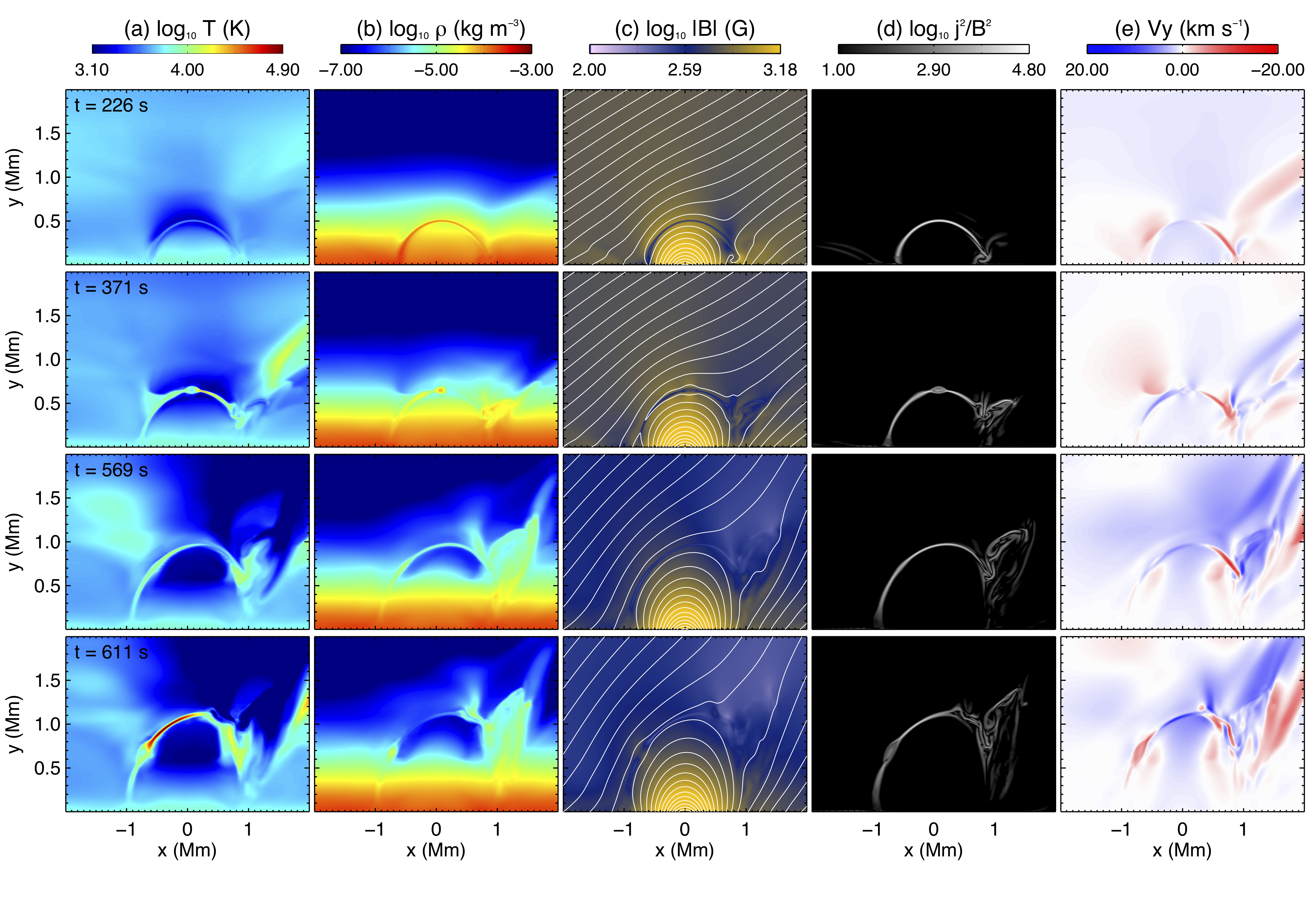}} 
\caption{The general view of Case 1 with the lowest resolution. The distributions of temperature (a), density (b), the total magnetic field strength (c), current density (d), and vertical velocity (e) at t=226, 375, 569, and 611 s are presented. The white lines in the third column outline the magnetic field lines. The initial background magnetic field is about $500$\,G.} \label{f2}
\end{figure*}

We started our simulations with fairly low resolutions to have a general  view of the magnetic reconnection processes. The highest AMR level in Case 1 is  1, giving a smallest grid size of $\sim 2.86$~km. The resolution in Case 1 is the lowest among all the Cases presented in this work, but it is still higher that those in previous 3D simulations in the low solar atmosphere \citep[e.g.,][]{Hansteen2019}.  Four snapshots of Case 1 is presented in Figure~\ref{f2}. The original data calculated from NIRVANA code is transformed to uniform IDL data. First, the new emerging flux rises from the bottom boundary to the computation domain. The emerging flux interacts with the existing overlaying magnetic field, and a current sheet forms along their separatrice. The current sheet is filled with dense low photosphere plasmas with a density that is larger than its ambient. The temperature inside the current sheet is not heated above 10,000 K. As the current sheet continues to rise to a higher location, plasmoids start to appear after it reaches 0.5 Mm above the solar surface. At the same time, the mass density in the current sheet decreases because the plasma is ejected to the exhaust region and falls down due to gravity. The density and temperature distributions along the current sheet become non-uniform after the plasmoids appear. Cores of these plasmoids fill with dense photosphere plasmas and the plasmas at the edges of the plasmoids are usually more tenuous (see Figs. 3), especially when they are growing bigger. Therefore, these regions with lighter plasmas are heated to higher temperature (> 20,000 K). Later on, the plasmoids are ejected to the exhaust regions and the current sheet continues to rise. In the end, an oblique current sheet extends from 0 to 1.2 Mm in height and the mass density at the higher part of the current sheet is much less than that at the lower part. Plasma at higher heights of the current sheet is heated to $\sim10^{5}$ K, while the temperature at lower heights of the current sheet is below $ 10^{4}$ K. Such a final stage in the Case 1 is similar to the results in \citet{Hansteen2019}. In their 3D RMHD simulations, a vertical current sheet extends more than 2 Mm from photosphere to the low corona, and the cool ($< 10^{4}$ K) and hot ($\gtrsim10^{5}$ K) plasmas are located at the opposite ends of the current sheet. 

\begin{figure*} 
\centering {\includegraphics[width=\textwidth]{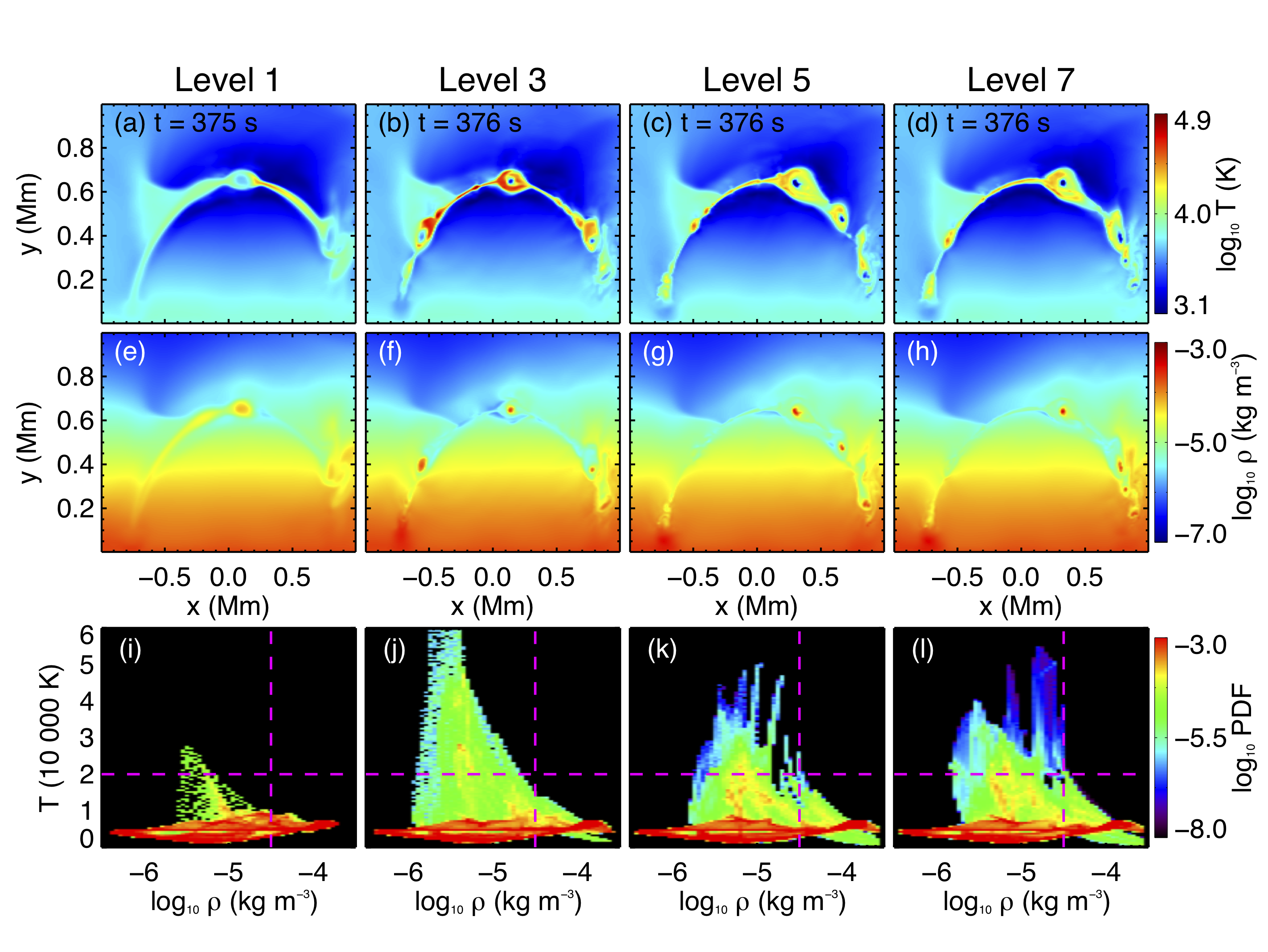}} 
\caption{The Comparisons among Cases 1, 2, 3 and 4 with different highest AMR levels. The highest AMR levels are 1, 3, 5 and 7 in these cases, respectively. The distributions of temperature (a)-(d), density (e)-(h), and joint probability distribution function of temperature and density (i)-(l) at around 376 s are presented. The vertical and horizontal purple dashed lines in panels (i)-(l) indicate the positions with the density of $10^{-4.5}$ kg\,m$^{-3}$ and the temperature of $2\times10^{4}$\,K.
} \label{f3}
\end{figure*}

\begin{figure*} 
\centering {\includegraphics[width=18cm]{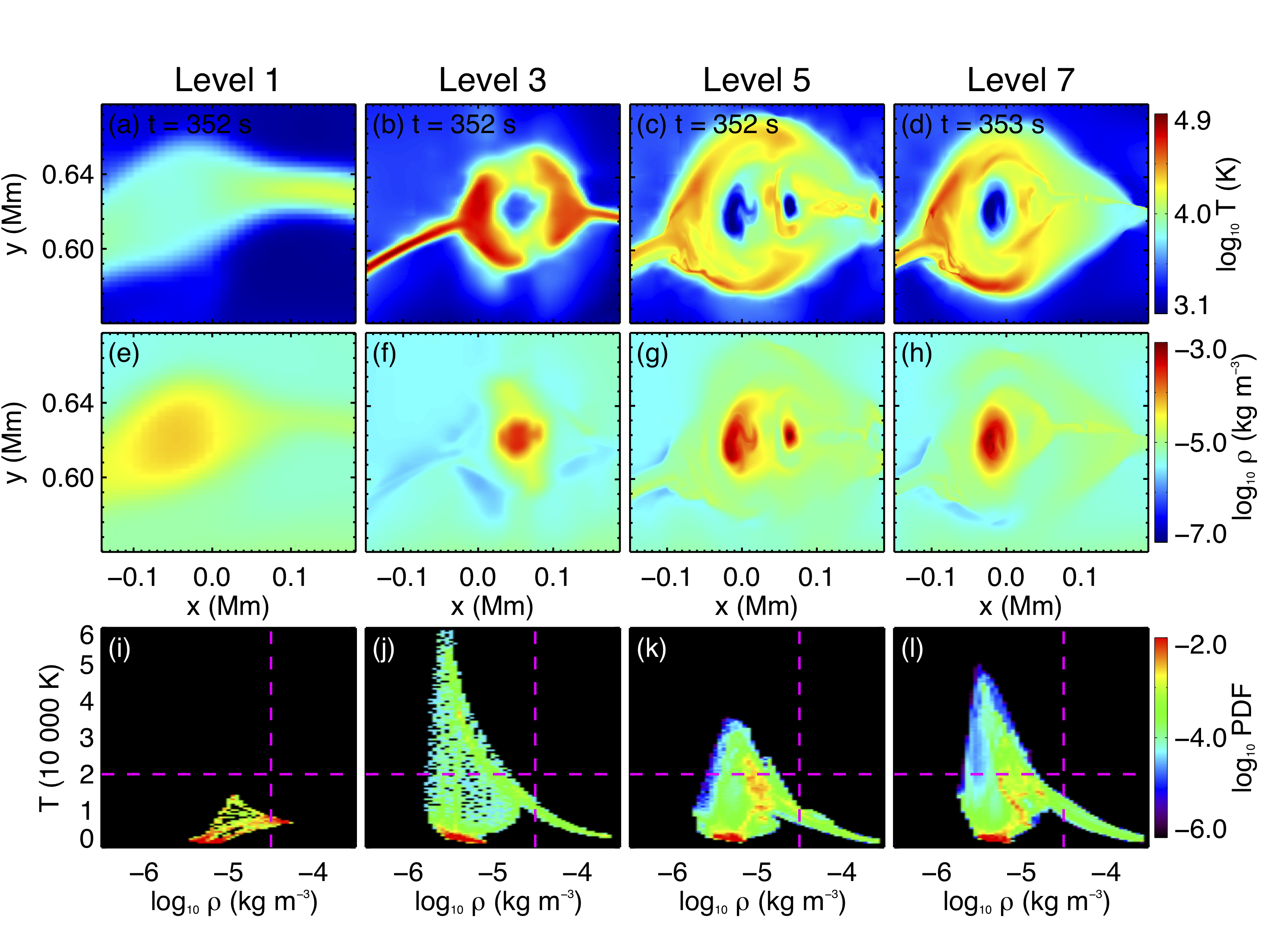}} 
\caption{The same as Figure~\ref{f3} but for smaller field of view in one plasmoid taken around 352 s.} \label{f4}
\end{figure*}

Based on the results of Case 1, we increased the highest AMR level step by step to investigate the effects of different computational spatial resolutions. In this work, we selected 4 cases with the highest AMR levels of 1, 3, 5, and 7, the comparisons among these cases are presented in Figure~\ref{f3} and Figure~\ref{f4}. Increasing resolutions in NIRVANA code makes the time step becoming smaller, the runs with higher resolutions are terminated earlier. However, the stages with plasmoid instability appearing in the reconnection region are well solved in all the cases. Figure~\ref{f3} presents the distributions of plasma temperature, density and the joint probability distribution function (PDF) of temperature and density at a time after plasmoids appear in the 4 cases. We zoomed into a small scale to show the fine structures in one plasmoid in  Figure~\ref{f4}. The Level one IDL data is used to plot Figure~\ref{f3}, but the original levels of IDL data are used in Figure~\ref{f4}. 

\begin{figure*}
    \centerline{\includegraphics[width=0.6\textwidth, clip=]{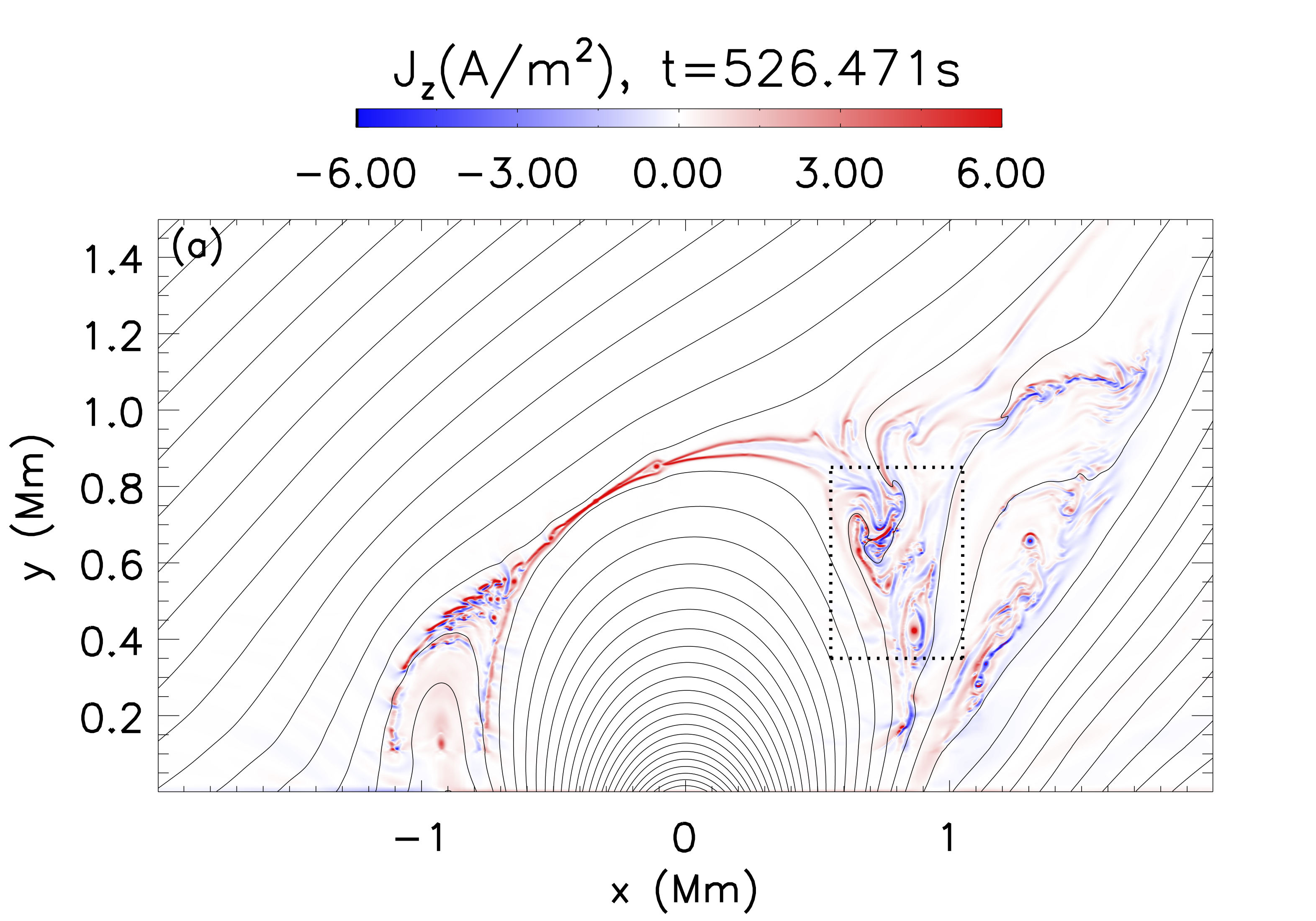}}
    \centerline{\includegraphics[width=0.4\textwidth, clip=]{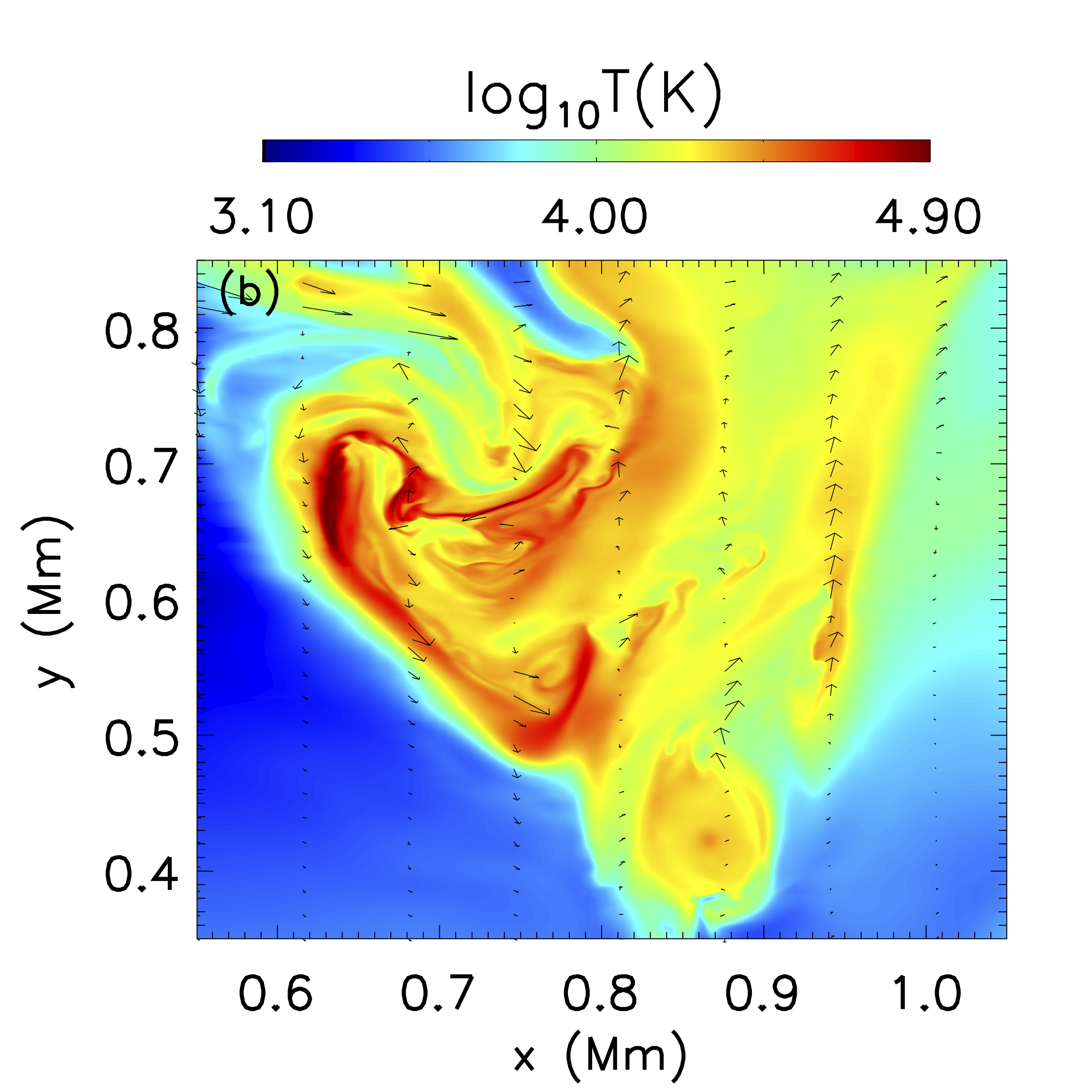}
                       \includegraphics[width=0.4\textwidth, clip=]{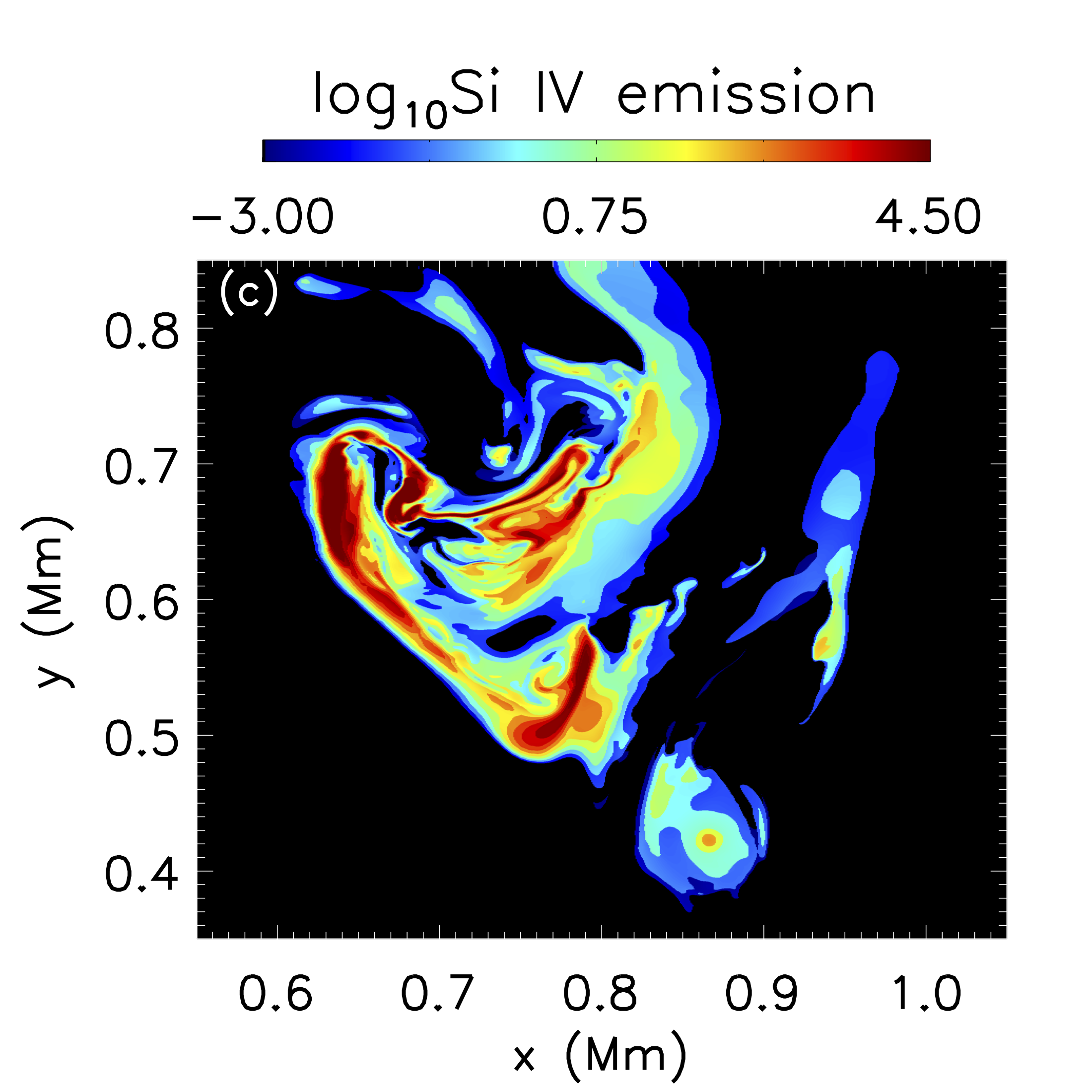}} 
  \caption{The distributions of the current density in z direction (a), temperature (b) and the synthetic Si~{\sc{iv}} 1394 {\AA} line emissions at each grid point (c) at a time during the later stage of the reconnection process in Case 3.  The domain in (b) and (c) is corresponding to the region in the dotted box in (a). The black arrows in (b) represent the total velocity.} \label{f5}
\end{figure*}

On can find that the locations and morphology of current sheets in all the cases are quite similar. However, the fine structures inside the current sheet are very different. Plasmoid instability appears in Case 1, but only the big plasmoids of first order are identified. These big plasmoids are very smooth, no turbulent structures appear inside them (See Figure~\ref{f3} and ~\ref{f4}). More plasmoids appear in Cases 3 and 4, and the turbulent fine structures inside the big plasmoids as shown in Figure~\ref{f4} are only clearly observed in these two cases. The numerical results in Case 3 with level 5 AMR are similar to those in Case 4 with level 7. The density and temperature distributions in the reconnection region in Case 1 are also not uniform, but the differences among different components in Case 3 and 4 are much larger and more obvious than those in Case 1, and more plasmas are heated to higher temperatures. Before the current sheet is emerged above the middle chromosphere, the maximum temperature in the reconnection region in Case 1 is always below 30, 000 K. Comparing Figure~\ref{f3} and ~\ref{f4}, one can find the maximum temperature in the main current sheet reaches above 60, 000 K in Case 2, and it is about 50, 000 K in Case 3 and 4. After the plasmoids are ejected out along the current sheet, they merge with the background plasmas and magnetic fields, both sides of the dome region become very turbulent and many fragment currents appear (See Figure~\ref{f5}), especially on the right hand side (See inside the black dotted box in Figure~\ref{f5}). Figure~\ref{f5} shows the results in Case 3,  such a turbulent reconnection out flow region is more clearly identified in the high resolution runs, the maximum temperature reaches about $80, 000$\,K in this turbulent region in Case 3. 

In the simulations, both physical diffusion and numerical diffusion contribute to heating. Taking ohmic heating as an example, the heating rate Q can be written as $Q=(\eta+\eta_{num} ) J^2=\eta_{eff} J^2$, where $\eta_{num}$ is numerical diffusivity and $\eta_{eff}$ is the total effective diffusivity. When the grid size becomes smaller, the current density increases and $\eta_{num}$ decreases.  In Case 1 and 2, the grids are coarse and the numerical diffusion dominates the heating. When the grid sizes become much smaller in Case 3 and 4, the numerical diffusion becomes much less and the physical diffusion starts to play a role. Though the current density in Case 3 and 4 is stronger than that in Case 2, the maximum temperature enhancement is less because of smaller global $\eta_{eff}$. As shown in Figure~\ref{f3} and Figure~\ref{f4}, the results in Case 3 are close to those in Case 4, which indicates that the simulation results in Case 3 and Case 4 are approaching the ones with realistic magnetic diffusion. 

%-----------------------------------------------------------------
\subsection{Effects of radiative cooling, heat conduction and ambipolar diffusion} \label{sec:terms}

\begin{figure*} 
\centering {\includegraphics[width=18cm]{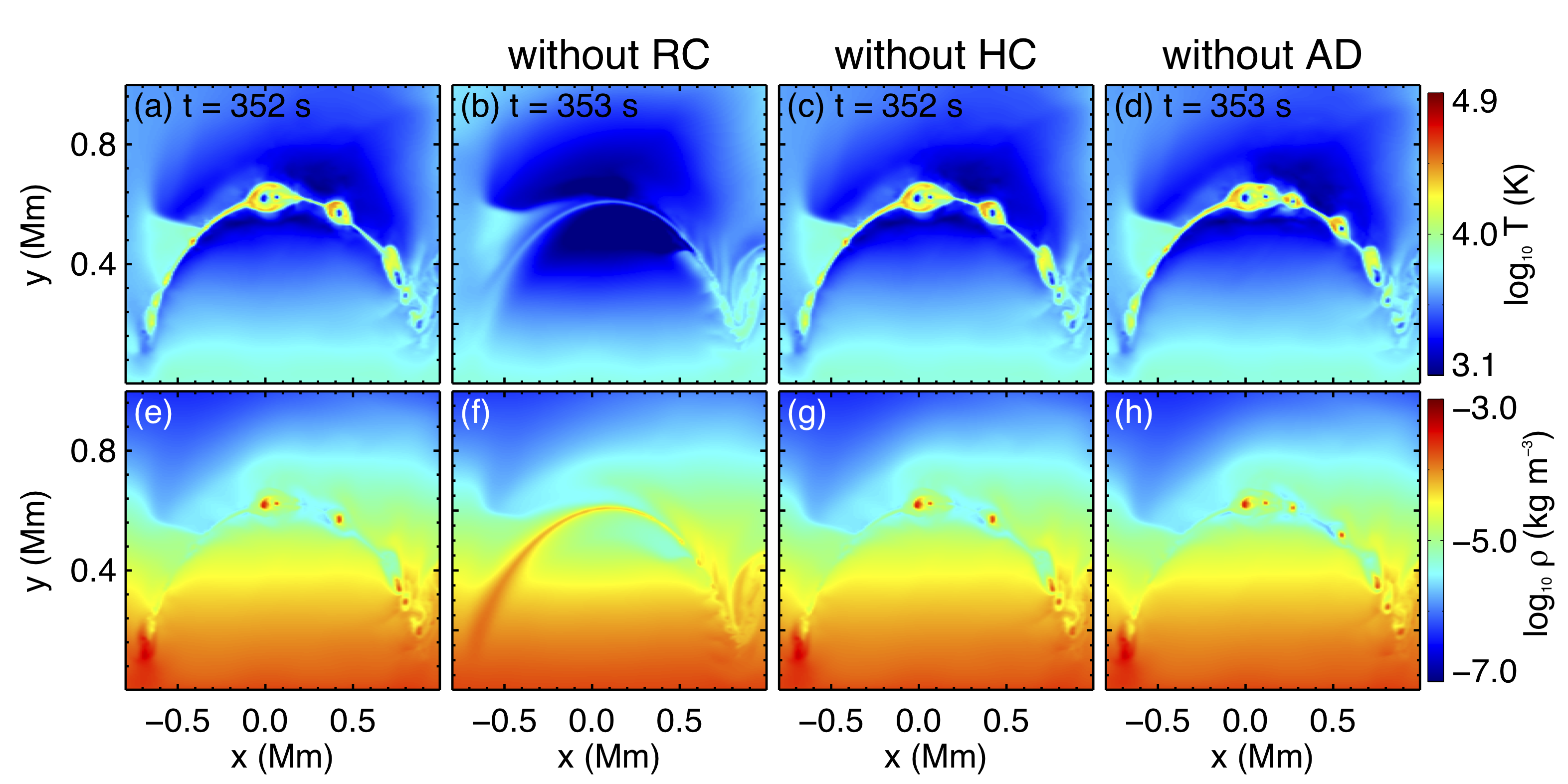}} 
\caption{The comparisons among cases (Cases 3, 3a, 3b and 3c) by turning off different terms. The distributions of temperature (a)-(d) and density (e)-(h) of cases 3, 3a (turning off radiative cooling), 3b (turning off heat conduction), and 3c (turning off ambipolar diffusion) are presented..} \label{f6}
\end{figure*}

\begin{figure*} 
\centering {\includegraphics[width=18cm]{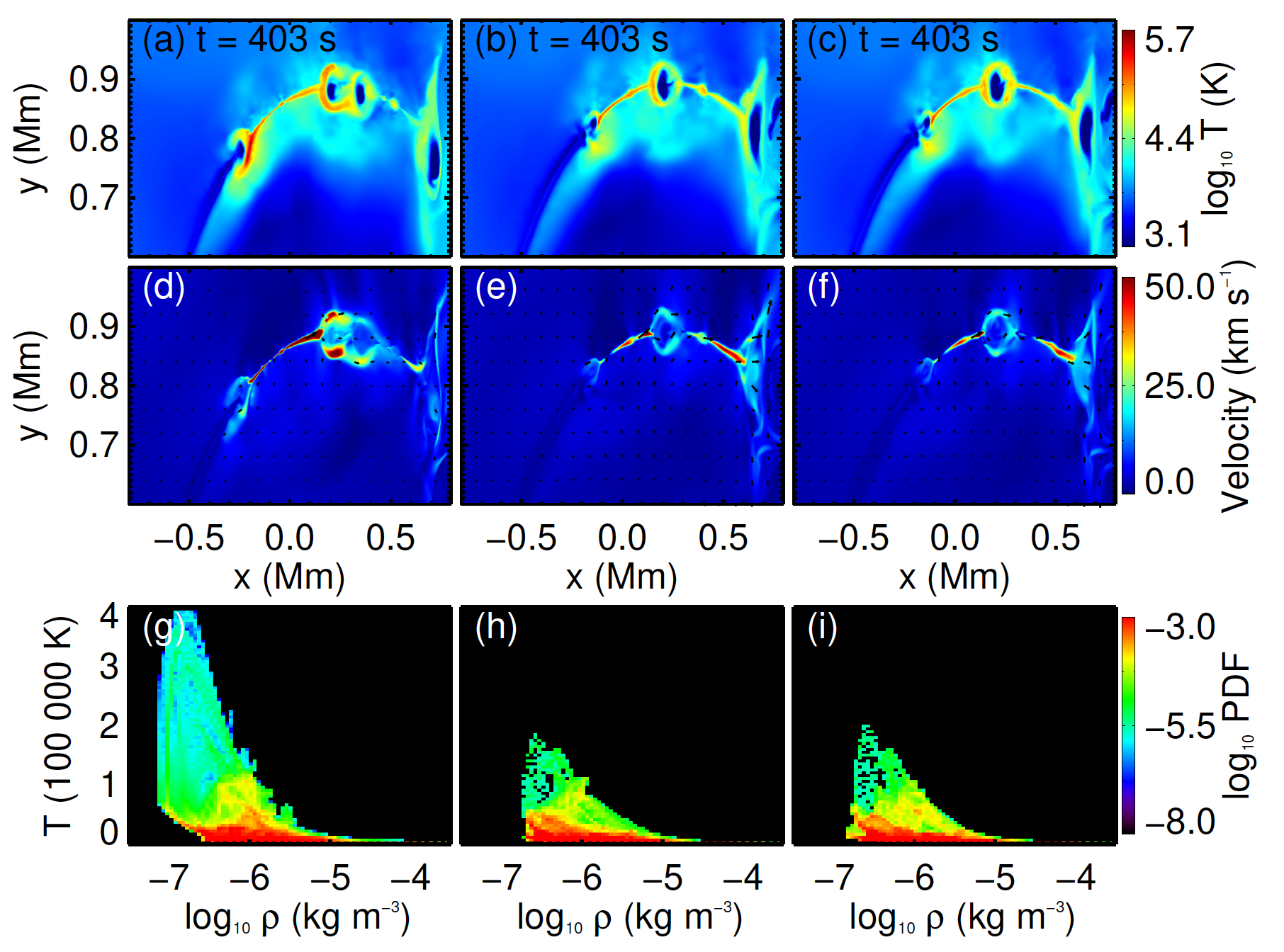}} 
\caption{The comparisons among three different cases. The distributions of temperature (a)-(c), total velocity in x-y plane ($\sqrt{v_x^2+v_y^2}$) (d)-(f), and the joint probability distribution function of temperature and density (g)-(i) for Case 5, Case 5a, and Case 5b are presented. (a), (d) and (g) are for Case 5, the highest AMR refinement level is 5 and ambipolar diffusion effect is not included. (b), (e) and (h) are for Case 5a, the highest AMR refinement level is 3 and ambipolar diffusion effect is not included. (c), (f) and (i) are for Case 5b, the highest AMR refinement level is 3 and ambipolar diffusion effect is included. These figures indicate that the effect of resolution on magnetic reconnection is stronger than ambipolar diffusion.} \label{f7}
\end{figure*}

\begin{figure*}
    \centerline{\includegraphics[width=0.4\textwidth, clip=]{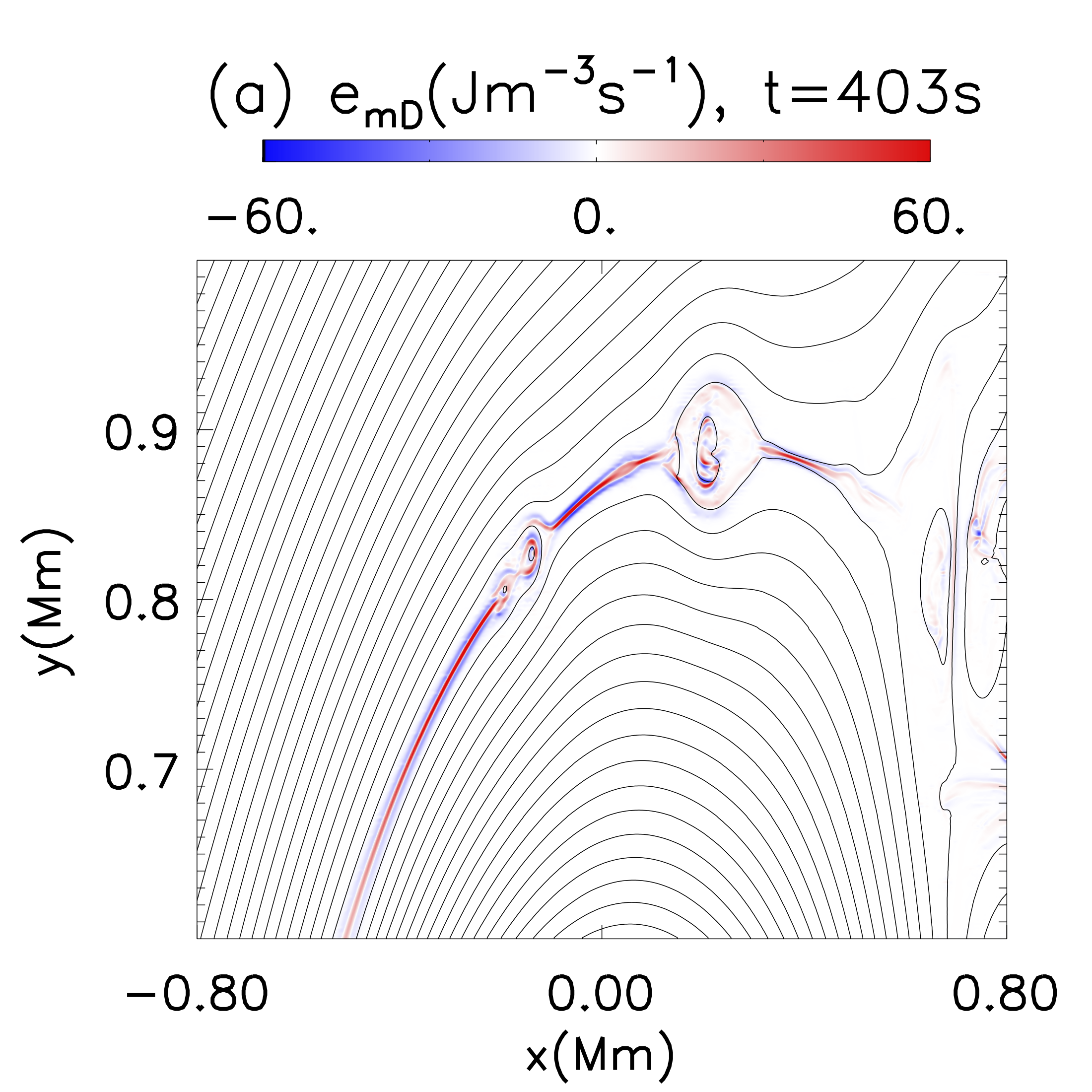}
                       \includegraphics[width=0.4\textwidth, clip=]{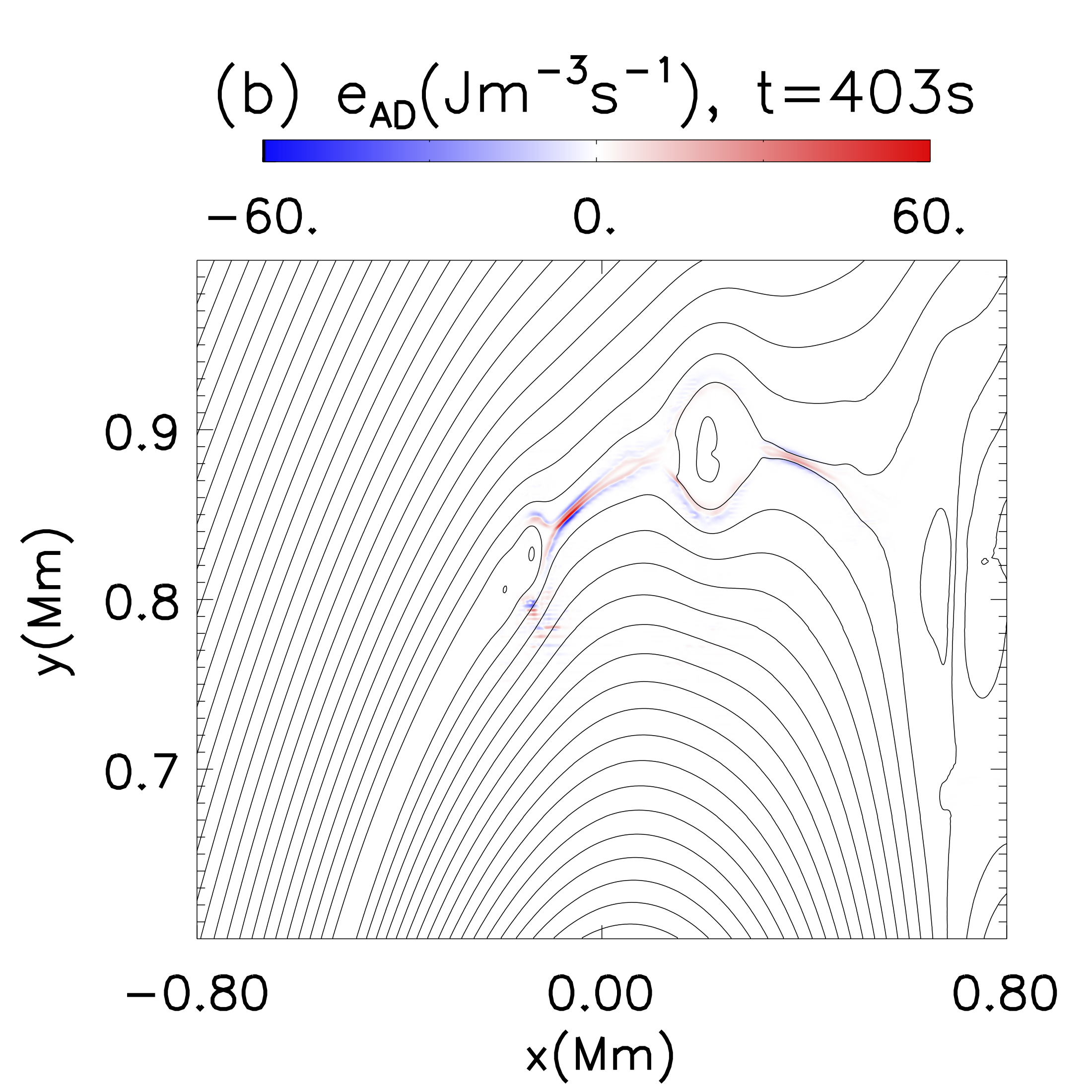}} 
  \caption{(a) shows the energy term $e_{mD}$ contributed by effective magnetic diffusion, (b) shows the energy term $e_{AD}$ contributed ambipolar diffusion. The data is from Case 5b. } \label{f8}
\end{figure*}

In order to understand the effects of radiative cooling in our simulations, we turned off radiative cooling and run the simulation again to obtain Case 3a. Similarly, we dropped heat conduction and ambipolar diffusion in Case 3b and 3c, respectively. The highest AMR levels in these Cases are 5, and we compared the results in Case 3a, 3b and 3c with Case 3 to study the three different effects on magnetic reconnection process. The current sheet in these cases is not emerged to a location higher than 0.7 Mm before they are terminated because of the extremely small time step. In order to check if the ambipolar diffusion can effect the magnetic reconnection process at the higher height, we also tried to include this effect in Case 5. However, including the ambipolar diffusion effect in Case 5 makes the run to be terminated before plasmoid instability appears. Therefore, we turned off the ambipolar diffusion effect in Case 5. Then, we performed another two simulations (Case 5a and 5b) by using less refinements. The highest AMR levels in Case 5a and 5b is 3. The ambipolar diffusion is turned off in Case 5a and included in Case 5b. These results are presented in Figure~\ref{f6}, ~\ref{f7} and ~\ref{f8}.
 
Comparing the results of Case 3 and 3a, one can find that radiative cooling plays important roles in such a magnetic reconnection process. Including radiative cooling makes plasmoids appear earlier (See Figure~\ref{f6}) and the dense cores of the plasmoids colder, which may significantly affect the synthesized spectral lines and images. Previous studies have suggested that radiative cooling has little effect on reconnection processes in the case with a strong guide field \citep[e.g.,][]{Uzdensky2011, Ni2015}. However, the guide field in our models is quite strong (500 G) and the radiative cooling still plays an important role in the reconnection process. The mass density in the current sheet region in our simulations is larger than previous simulations, which is probably one of the reasons to cause such a difference. Moreover, the reconnection process is driven by strong flux emergence process in this work, which is different from the previous current sheet studies with small initial perturbations  \citep[e.g.,][]{Ni2015}. We also find that the upper boundary is unrealistically heated to high temperatures in the case by excluding radiative cooling. Therefore, including radiative cooling is very important for numerical studies of activities in the low solar atmosphere. In this work, we have used the radiative cooling model proposed in the previous works by \cite{Gan1990}, in which the authors have derived their model based on the detailed non-LTE calculations. Using the plasma parameters from the low solar atmosphere model \citep{Vernazza1981}, they found that the radiative loss calculated from their model is close to that from non-LTE calculations. Therefore, this model can give a reasonable approximation for the radiative cooling in the photosphere and chromosphere. However, the more realistic models in further studies are still necessary. Comparing Figure~\ref{f6}a and ~\ref{f6}c, one can find that heat conduction does not play a role in the main current sheet region. The maximum temperature reaches from several thousands K to tens of thousands K, which  might make the heat conduction along the magnetic field line become significant. However, the plasmoids with closed magnetic fields trap the heating within the plasmoids, and the heat conduction loses efficacy.

The previous studies show that ambipolar diffusion can be very important in heating the chromosphere, amplify and transport the magnetic tension \citep[e.g.,][]{Khomenko2012, Martinez-Sykora2017} in the low solar atmosphere. Our simulation results indicate that ambipolar diffusion does not significantly change the reconnection process and local plasma heating in the low chromosphere. Comparing  Figure~\ref{f6}a and ~\ref{f6}d, one can find that there are small differences in details between Case 3 and 3c. Figure~\ref{f7} also shows that the ambipolar diffusion effect makes the maximum temperature in Case 5c to be slightly higher than those in Case 5b. We have calculated the two terms $e_{mD}$ and $e_{AD}$ on the right-hand side of the energy equation at $t=403s$ in Case 5c with ambipolar diffusion, where $e_{mD}=\nabla \cdot \left [\frac{\eta_{eff} }{\mu _{0}}\mathbf{B}\times \left ( \nabla \times \mathbf{B} \right )\right ]$ and $e_{AD}=-\nabla \cdot \left [ \frac{1 }{\mu _{0}}\mathbf{B}\times\mathbf{E}_{AD}\right ]$. The distributions of the  two terms contributed by magnetic diffusion and ambipolar diffusion at this moment are presented in Figure~\ref{f8}. One can find that the maximum values of $e_{AD}$  and $e_{mD}$ are about the same magnitude. As shown in Figure~\ref{f9}c and ~\ref{f9}d, the plasma density inside and around the current sheet region is $\gtrsim10^{20}$\,m$^{-3}$, such a high plasma density makes the maximum ambipolar diffusion is comparing with the maximum effective diffusion in our simulations. However, the areas with strong $e_{mD}$ are much larger than the areas with strong $e_{AD}$. The heating contributed by compression processes is also very strong. Therefore, the heating contributed by ambipolar diffusion is much less efficient comparing with the other processes. As we zoom into the regions with strong $e_{AD}$  and $e_{mD}$, we also find that the locations with strong $e_{AD}$ are always staggered with the locations with strong $e_{mD}$. We should point out that the ionization fraction $Y_i$ is fixed in our simulations. Since the realistic ionization fraction $Y_i$ should increase with temperature and the ambipolar diffusion $\eta_{AD}$ decreases with increasing $Y_i$ and T, heating caused by ambipolar diffusion is possibly overestimated in our simulations. The ambipolar diffusion effect should become even less efficient than that we have shown in this work if we consider the time dependent ionization fraction. Ambipolar diffusion can possibly play important roles on the magnetic reconnection process above the middle chromosphere, where the plasma density is much lower. In the recent work of \cite{Nobrega2020}, the authors found that the ambipolar diffusion does not significantly change the amount of emerged magnetic flux when the non-equilibrium ionization and recombination and molecule formation of hydrogen is included, but it can efficiently heat the shock structures above the middle chromosphere.

%-----------------------------------------------------------------
\subsection{Synthetic Si~{\sc{iv}} line profiles} \label{sec:spec}

\begin{figure*}
\centering{\includegraphics[width=18cm]{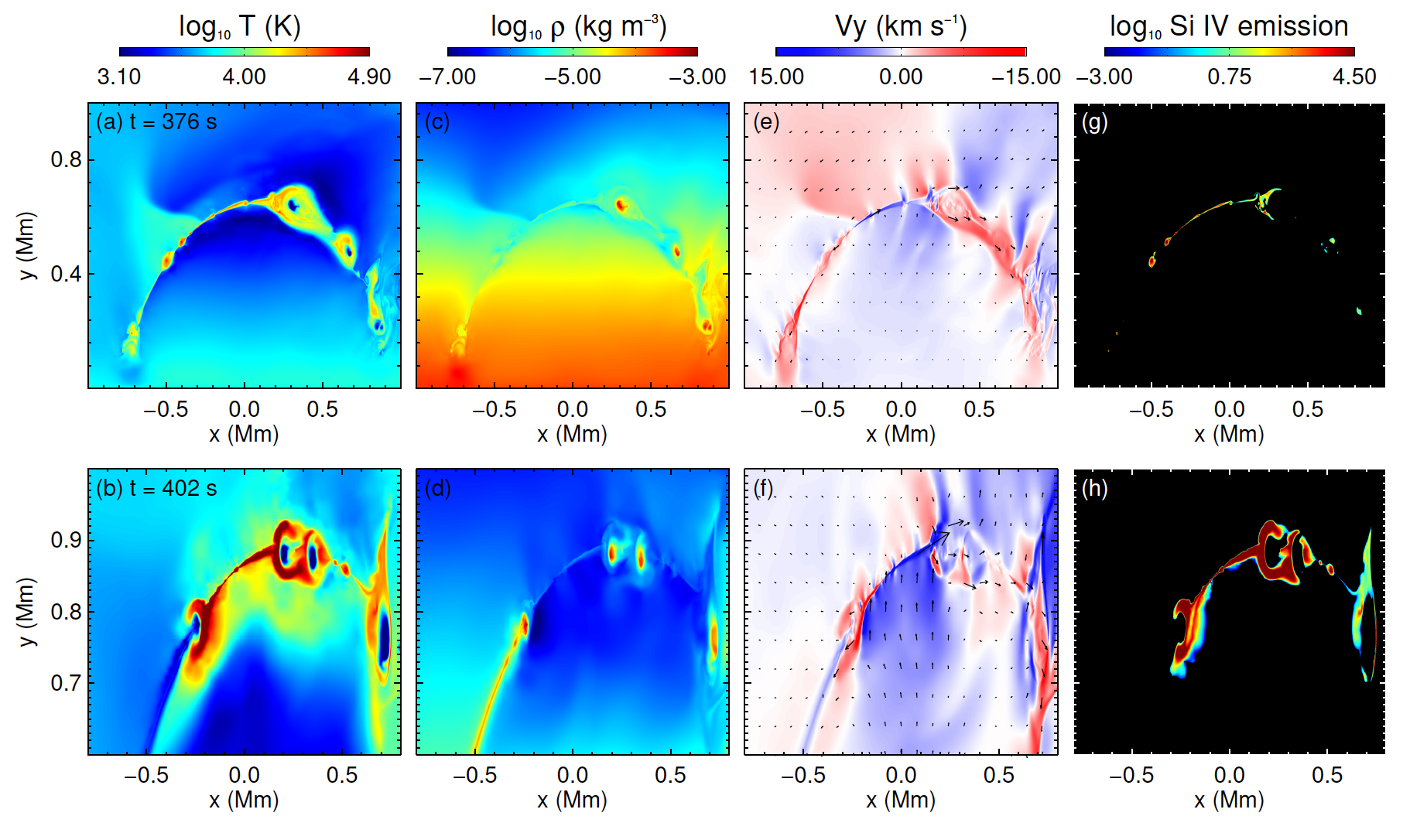}}
\caption{The comparisons of the main current sheet region in Case 3 (top panels) and 5 (bottom panels) at the times when the plasmoid instability is well developed.  The distributions of temperature (a)-(b), density (c)-(d), vertical velocity $V_y$ (d)-(f), the synthetic Si~{\sc{iv}} 1394 {\AA} line emissions at each grid point (g)-(h) are presented. The black arrows in (e) and (f) represent the total velocity. The background magnetic fields are 500 G and 600 G in Case 3 and Case 5 respectively. Case 5 has larger emerging flux than Case 3, the heat conduction and ambipolar diffusion effects are not included in Case 5. Panels (g) and (h) are shown in arbitrary units.} \label{f9}
\end{figure*}

\begin{figure*}
    \centerline{\includegraphics[width=0.4\textwidth, clip=]{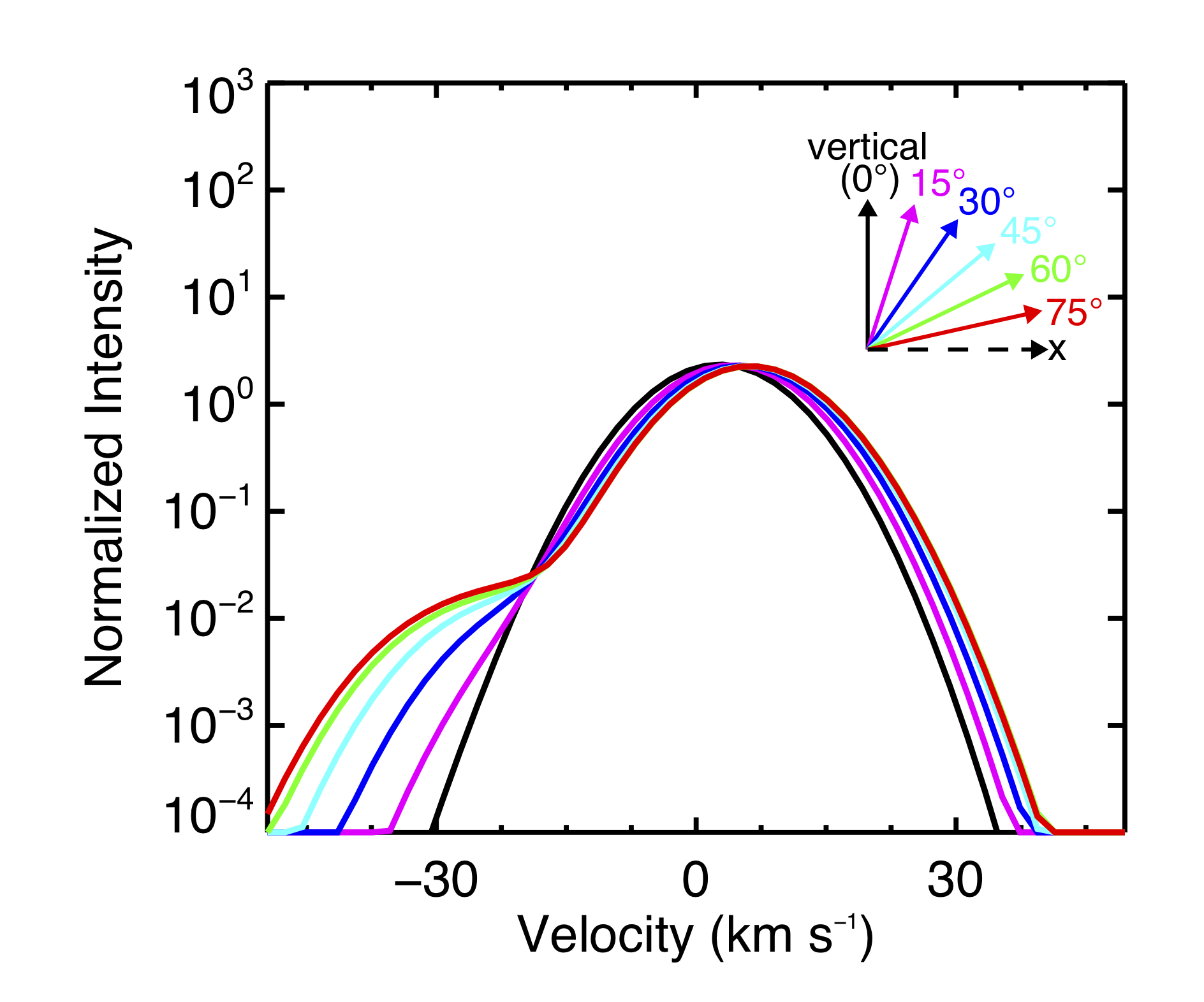}
                       \includegraphics[width=0.4\textwidth, clip=]{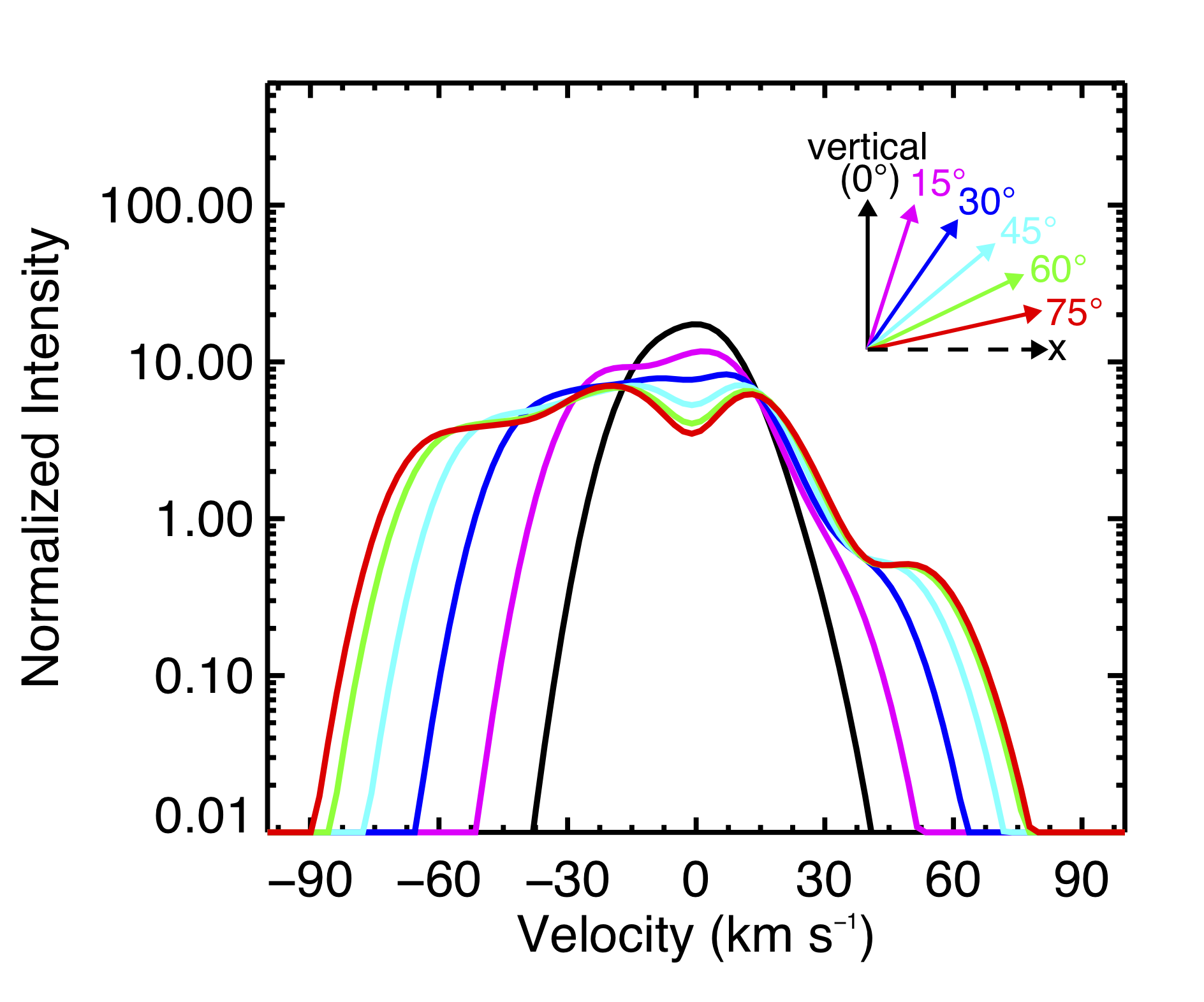}} 
  \caption{The Synthetic Si~{\sc{iv}} 1394 {\AA} line profiles in Case 3 and Case 5 taken along different lines of sight are presented in the left panel and right panel respectively. The Si~{\sc{iv}} line profiles are integral over the whole regions shown in the Figure. 9.  Black, purple, blue, cyan, green, and red curves represent the line profiles taken along different lines of sight, the angles between the lines of sight and the vertical direction (y-direction) are  $0^{\circ}$, $15^{\circ}$, $30^{\circ}$, $45^{\circ}$, $60^{\circ}$, $75^{\circ}$, respectively. The dashed arrow represents the x-axis of the computation domain, and the solid arrows with different colors indicate the directions of lines of sight taking for calculating the Si IV line profiles shown with the same colors. } \label{f10}
\end{figure*}

We calculated the emissions and line profiles of the Si~{\sc{iv}} 1394 {\AA} line of Case 3 and 5 based on the method described in \citet{Peter2006} using the atomic data package CHIANTI \citep[version 9.0;][]{Dere2019}. Level 5 IDL data were used for calculations. First, we calculated the emissivity of the Si~{\sc{iv}} 1394 {\AA} line at each grid point. The emissivity $\varepsilon$ depends on density and temperature and can be written as:
\begin{equation}
\varepsilon=G\left( n_{e},T \right) n_{e}^{2}
\end{equation}
where $n_{e}$ is the electron number density and $G\left( n_{e},T \right)$ is the contribution function which can be calculated from CHIANTI. Assuming that the line profile at each grid point has a thermal width of $w_{th}=\sqrt{2k_{B}T/m_{i}}$, we calculated the spectral line profiles at each grid point with the unit of Doppler velocity:
\begin{equation}
I\left(v\right)=\frac{\varepsilon}{\sqrt{\pi}w_{th}}\exp{ \left [-\frac{\left(v-v_{0} \right)^{2}}{w_{th}^{2}} \right]}
\end{equation}
where $v_{0}$ is the component of velocity along the line of sight. Finally, we integrated the spectral line profiles over the whole regions shown in Figure~\ref{f9} to obtain the total spectrum shown in  Figure~\ref{f10}. 

In Case 3, we noticed that the hydrogen density within the current sheet is as large as $10^{15}$\,cm$^{-3}$, and the emissions are underestimated because the Si~{\sc{iv}} has near Saha-Boltzmann opacity and its formation temperature drops to $10^{4} \sim 2\times10^{4}$\,K at such a high density \citep{Rutten2016}. However, there are still obvious Si~{\sc{iv}} emissions in the current sheet by using such a calculation method based on the optically thin assumption. The newly formed turbulent region also has strong emissions (See Figure~\ref{f5}), and this region has a diameter of about 0.4 Mm. The stronger initial background magnetic field and emerged magnetic flux have been applied in the model to simulate Case 5. Comparing the results shown in Figure~\ref{f9}, one can find that the main current with multiple plasmoids is emerged to a higher location in Case 5, the plasmas are heated to much higher temperatures and the maximum velocity is also higher in Case 5. The emissions in the reconnection region in Case 5 are much stronger than those in Case 3 (See Figure~\ref{f9}g and ~\ref{f9}h). 

When the line of sight (LOS) is perpendicular to the direction of the reconnection outflow, the synthesized Si~{\sc{iv}} spectral profile is narrow and close to a single Gaussian profile (See the black solid lines in Figure~\ref{f10}). As the LOS deviates from the direction perpendicular to the reconnection outflow, the doppler effect appears, and the spectral line profile broadens and is no longer a single Gaussian profile. The multiple turbulent plasmoids along the LOS is the reason to cause the non-Gaussian line profile. However, the line width in Case 3 is only slightly broadened when the LOS passes through multiple plasmoids (See Figure~\ref{f10}a). We find that the line width of the synthesized Si~{\sc{iv}} spectral line profile strongly depends on the maximum velocity along the LOS, which depends on both the local Alfv\'en velocity near the reconnection inflow region and the angle between the LOS and the reconnection outflow. Assuming the strength of the reconnection magnetic field ranges from $200$ to $2,000$ G, the corresponding plasma density is then between $4\times 10^{14} \sim 1.6\times10^{21}$\,m$^{-3}$ in order to obtain a local Alfv\'en velocity of $100$\,km s$^{-1}$. Therefore, the reconnection region for generating the UV emissions with broad Si~{\sc{iv}} line width over $100$\,km s$^{-1}$ generally should locate above the solar TMR. Our simulations in Case 5 with reconnection magnetic fields of $\sim600$\,G show that the high-speed plasma that accounts for the broad Si IV line is located around 0.85 Mm above the solar surface (See Figure~\ref{f9}b, \ref{f9}d, \ref{f9}f, \ref{f9}h). When the LOS is parallel to the direction of the maximum reconnection outflow velocity, the blue wing of the line profiles extends to about $100$\,km s$^{-1}$. If the reconnection region is extended to a higher position or the reconnection magnetic field is stronger than that in Case 5, a broader line width can be generated.
 
UV bursts with small line widths have also been reported \cite{Hou2016}, and their formation mechanisms are still unclear. Our simulations show that the narrow Si IV line profiles (See Figure~\ref{f10}a) can easily form in the reconnection region around and above the solar TMR, as long as the reconnection magnetic fields are strong enough to make the plasmas to be heated above $20,000$\,K. Comparing with the UV bursts with broad line widths, the UV bursts with narrow line widths can be formed in a reconnection region at a lower height or with weaker reconnection magnetic fields (Figure~\ref{f10}a), or the LOS direction and the reconnection outflow direction is almost perpendicular to each other (Figure~\ref{f10}b).

%--------------------------------------------------------------------------------------------------------------------------%

\section{Summaries, Conclusions and Discussions}

In this work, magnetic reconnection between the emerged and back ground magnetic fields has been simulated to study the UV bursts which relating with EBs. The strength of the reconnection magnetic fields with opposite directions is about several hundred Gauss after the current sheet reaches 0.5 Mm above the solar surface. The stratified atmosphere is close to the C7 model \citep{Avrett2008}, the initial temperature varies with height and a temperature minimum region is included, the plasma density changes more than five order of magnitude from the bottom to  2\,Mm above the solar surface. The partial ionization effect including ambipolar diffusion is considered in simulations. An empirical radiative cooling model based on observations is applied, and an anisotropic heat conduction is also included in our simulations. The simulations are performed by using different ARM levels, which results in different resolutions in the current sheet region in different cases.  

Our simulations can explain aspects of the formation of UV bursts. The hot tenuous plasmas and the cold dense plasmas carried upward from the photosphere are mixed up during plasmoid instabilities inside the reconnection region. The ejected plasmoids are then further merged into the background magnetic fields and plasmas, which causes the formation of the new turbulent region with many fragment currents. The temperatures in the plasmoids and the new turbulent region can range from several thousands K to about $100,000$ K, which could explain the different emission and absorption lines observed in a UV burst. The multiple turbulent structures along the LOS is the reason to cause the non-Gaussian line profile of Si~{\sc{iv}}. However, the line width of the synthesized Si~{\sc{iv}} spectral line profile strongly depends on the maximum velocity along the LOS. The roundish shape of the plasmoids and the new turbulent region also match well with the shape of UV bursts from observations \citep[e.g.,][]{Peter2014, Tian2016, Chen2019b}, which is different from the narrow elongated shape from the 3D simulations. Our simulations can also explain the association between UV bursts and EBs. The hot plasmas ($> 20,000$\,K) inside the current sheet appear during the later stage of the reconnection process, which explains the fact that UV bursts are usually seen after their EB counterparts \citep{Ortiz2020}. The hot turbulent structures with strong Si IV emissions mainly appear above and around the solar TMR, and the cold plasma extends to much lower locations near the bottom of the current sheet at the solar surface (See Figure ~\ref{f5} and ~\ref{f9}). These phenomena are consistent with the observations that UV bursts have a tendency to locate at higher heights of corresponding EBs and that EBs have partial overlap with their corresponding UV bursts in space. 

The previous 2.5D simulations have also shown the formations of plasmoids and the well sythesized Si~{\sc{iv}} spectral in a stratisfied solar atmosphere with flux emergency \citep[e.g.,][]{Nobrega2016, Rouppe2017}. However, we should point out that plasmoids are formed above the up chromosphere and the density in UV bursts is $\lesssim 10^{18}$\,m$^{-3}$ in those previous simulations \citep{Nobrega2016, Rouppe2017}. The surface flows at the bottom boundary can also self-consistently drive the formation of the reconnection region of UV burtsts \citep{Peter2019}. Instead of using a realistic density stratification, the cases with different plasma densities are tested to study the effects of plasma $\beta$ on generating UV bursts in \cite{Peter2019}. The reconnection magnetic field in \cite{Peter2019} is about $50$\,G, the plasma with a density $>10^{19}$\,m$^{-3}$ is not heated above $20,000$ K in their simulations. The density of UV bursts is suggested $\gtrsim 10^{19}$\,m$^{-3}$ from observations \citep{Young2018}, it could be even higher for those connecting with EBs \citep[e.g.,][]{Tian2016}. When the plasma $\beta$ is the same, the temperature increase is much less in the plasma with a higher density in the lower atmosphere \citep{Ni2018c}. Therefore, it is more difficult for plasmas to be heated to high temperatures above $20,000$\, K in the lower atmosphere. In this work, plasmoids are formed below the middle chromosphere, the density in UV bursts is about 2 to 3 orders of magnitude higher and the reconnection magnetic fields are also stronger than those in the previous models.  Furthermore, the cool cores of the plasmoids and the multi-thermal turbulent fine structures inside the plasmoids and the reconnection outflow regions were not clearly shown in those previous simulations. 

 In the 3D simulations of \cite{Hansteen2019}, the part of the long current sheet with Si~{\sc{iv}} emissions downwardly extends to a position around the middle chromosphere. However, the plasmoid like structures are not shown in their work. The hot and cool parts are basically located at opposite ends of a long current sheet \citep{Hansteen2019}. They further proposed \cite{Ortiz2020} that the coexisting EBs and UV bursts would be part of the same reconnection system, but happening far apart vertically, which is different from the model in this work. Though our results also show that the cold EB part extends to a lower height, the cold and hot plasmas could be concentrated in one plasmoid or the turbulent outflow region at about the same height. In the recent work by \citep[e.g.,][]{Priest2018, Syntelis2019, Syntelis2020}, they have studied magnetic reconnection driven by photospheric flux cancellation. They pointed out that the hot and cool outflows produced without time difference and spatial offset are also found in a reconnection process without plasmoid like structures. 
 
 The more realistic radiative transfer process has  been well included in some of the previous simulations \citep[e.g.,][]{Nobrega2016, Rouppe2017, Hansteen2019} for studying UV bursts. However, we have considered the partial ionization effects and included the more realistic magnetic diffusion. 

The simulations in this work have coupled the macroscopic scale of several Mm and the microscopic scale down to the ion inertial length ($\sim45$\,m), both the observational features and the hidden small scale physics are revealed. The results indicate that the high resolution simulation including the realistic diffusivity in the stratified low solar atmosphere is essential for better understanding the magnetic reconnection mechanisms and plasma heating during the formation process of the UV bursts. Obviously, low resolutions inevitably result in the large numerical diffusivity and the low Lundquist number. The large numerical diffusivity might result in the unrealistic plasma heating in the reconnection region (as shown in Cases 1 and 2). The plasmoid instability will not appear if the Lundquist number is too low \citep[e.g.,][]{Bhattacharjee2009, Ni2013, Leake2012}, and the mixed hot tenuous and cold dense plasmas in the plasmoids and the turbulent reconnection outflow region will be smoothed out. This work qualitatively reveals the multi-thermal turbulent structures in the reconnection region of a UV burst. However, the non-equilibrium ionization-recombination and more realistic radiative transfer in the future simulations are important to find out the exact temperatures of the plasmas in these events. The obvious flux cancellations at the solar surface usually appear during the formation process of EBs and UV bursts \citep[e.g.,][]{Peter2014, Tian2016, Chen2019b}, which should also be included and discussed in the future more realistic 3D simulations.
 
\begin{acknowledgements}
We thank Professor Rony Keppens for helpful discussions. This research is supported by the Strategic Priority Research Program of CAS with grants XDA17040507 and QYZDJ-SSWSLH012; the NSFC Grants 11973083; the Youth Innovation Promotion Association CAS 2017; the Applied Basic Research of Yunnan Province in China Grant 2018FB009; the Yunnan Ten-Thousand Talents Plan-Young top talents; the project of the Group for Innovation of Yunnan Province grant 2018HC023; the YunnanTen-Thousand Talents Plan-Yunling Scholar Project; the Special Program for Applied Research on Super Computation of the NSFC-Guangdong Joint Fund (nsfc2015-460, nsfc2015-463, the second phase); the Max Planck Partner Group program.Y.C. is supported by the China Scholarship Council for his stay in Germany. Y.J. and H.T. are supported by NSFC grants 11825301, 11790304(11790300), and the Max Planck Partner Group program.
\end{acknowledgements}


\end{document}